\documentclass[tighten,twocolumn,apj]{likeapj}
\usepackage{apjfonts}
\shortauthors{Hall \& Fu}

\received{December 6, 2023}
\revised{April 22, 2024}
\accepted{May 1, 2024}

\newcommand{\kms}{{km\,s$^{-1}$}}
\newcommand{\ergps}{{erg\,s$^{-1}$}}
\newcommand{\msun}{$M_\odot$}
\newcommand{\msunyr}{$M_\odot\,{\rm yr}^{-1}$}
\newcommand{\lsun}{$L_\odot$}
\newcommand{\um}{$\mu$m}

\newcommand{\pcm}{cm$^{-2}$}

\newcommand{\sbunit}{erg\,s$^{-1}$\,cm$^{-2}$\,arcsec$^{-2}$ }
\newcommand{\surf}{cgs}

\newcommand{\logNHI}{\log N_{\rm HI}}

\newcommand{\Ha}{H$\alpha$}

\newcommand{\Lya}{Ly$\alpha$}
\newcommand{\Lyb}{Ly$\beta$}

\newcommand{\HI}{H\,{\sc i}}

\newcommand{\HeII}{He\,{\sc ii}}
\newcommand{\OII}{[O\,{\sc ii}]}

\newcommand{\cothree}{CO\,(3$-$2)}

\newcommand{\system}{GAMA\,J0913$-$0107} 
\newcommand{\SMGfull}{ALMA\,J091339.55$-$010656.4} 
\newcommand{\bgQSOfull}{SDSS\,J091338.97$-$010704.6} 
\newcommand{\fgQSOfull}{SDSS\,J091338.30$-$010708.6} 
\newcommand{\Chost}{Comp b} 
\newcommand{\bgQSO}{QSO1} 
\newcommand{\fgQSO}{QSO2} 
\newcommand{\objone}{Obj 1} 
\newcommand{\objtwo}{Obj 2} 
\newcommand{\objthree}{Obj 3} 
\newcommand{\objfour}{Obj 4} 

\newcommand{\mycomment}[1]{}
\usepackage{amsmath}

\defcitealias{Fu21}{Paper I}

\begin{document}

\title{\HI\ \Lya\ Emission from a Metal-Poor Cool Stream Fueling an Early Dusty Starburst}
\author{
Kevin~Hall and Hai~Fu
}
\affiliation{Department of Physics \& Astronomy, University of Iowa, Iowa City, IA 52242}

\begin{abstract}

The \system\ system is a rare conjunction of a submillimeter galaxy (SMG) at $z \approx 2.7$ and two background QSOs with projected separations $<$200\,kpc. Previous high-resolution QSO absorption-line spectroscopy has revealed high \HI\ column density, extremely metal-poor ($\sim 1\%$ solar) gas streams in the circumgalactic medium of the SMG. Here we present deep optical integral-field spectroscopy of the system with the Keck Cosmic Web Imager (KCWI). Reaching a $2\sigma$ surface brightness (SB) limit $\approx 10^{-19}$\,\sbunit\ with $\sim$2\,hrs of integration time, we detect a filamentary \Lya\ nebula stretching $\sim$180\,kpc from the SMG intercepting both QSO sightlines. This \Lya\ filament may correspond to the same cool gas stream penetrating through the hot halo seen in the absorption. In contrast to \Lya\ nebulae around QSOs, there is no obvious local source for photoionization due to the massive dust content. While uncertain, we consider the possibility that the nebula is ionized by shocks induced by the infall, obscured star formation, and/or a boosted UV background. The SMG-QSOs conjunction multiplied the efficiency of the KCWI observations, allowing a direct comparison of \Lya\ nebulae in two distinct environments. We find that the nebula around the QSOs are much brighter and show steeper surface brightness profiles than the SMG nebula. This is consistent with the additional photoionization and \Lya\ scattering provided by the QSOs. While illustrating the challenges of detecting \Lya\ nebulae around SMGs, our work also demonstrates that important insights can be gained from comparative studies of high-$z$ \Lya\ nebulae.

\end{abstract}


\section{Introduction} \label{sec:intro}

The study of galaxy formation and evolution remains an active field of research in astrophysics, encompassing a multitude of unanswered questions. One such question is how galaxies acquire the cold and dense gas to grow their stellar content. As dark matter (DM) halos grow, gas gradually accumulates within, marking the inception of galaxy formation from primordial density fluctuations. Above a mass threshold of $M_{\rm shock}$ $(\sim 10^{12} $ \msun), analytical and numerical models postulated that the inflow of gas develop shocks in the vicinity of the virial radius of the halo, heating the gas to the virial temperature ($T_{\rm vir} \gtrsim 10^6$ K) \citep[e.g.,][]{Rees77, Silk77, White78, Birnboim03}. Radiative cooling of such hot gaseous halos is so slow that it cannot provide enough cold gas for the galaxies in these halos to grow at significant rates. However, more recent numerical simulations showed that this picture may be incomplete. For massive halos ($M_{\rm DM} > M_{\rm shock}$) at $z \gtrsim 1.5$, although diffuse shock-heated hot gas still dominates the circumgalactic medium (CGM), the cosmic web connecting massive halos may be resilient enough to penetrate through the hot gas halo and supply chemically pristine cool gas ($\sim 10^4$\,K) from the intergalactic medium (IGM) to the central galaxy. The newly predicted picture is multiple long filaments of cool gas streams embedded in hot X-ray emitting halos \citep{Keres05,Keres09,Dekel06,Dekel09,Faucher-Giguere11,Martin19}.

The presence of cool gas streams in hot halos could be a pivotal mechanism in elucidating the existence of dusty galaxies exhibiting exceptionally elevated star formation rates (SFRs) at $z > 2$, such as the submillimeter-bright galaxies \citep[SMGs;][]{Smail97,Barger98,Blain02}. Despite being at significant cosmological distances, the SMGs are the brightest sources selected between 850\,\um\ and 2\,mm thanks to their huge infrared luminosity and the negative $K$-correction in the Rayleigh–Jeans tail. Their pronounced infrared luminosities can be ascribed to the substantial presence of dust obscuring intense star formation (SFR > 500\,\msunyr). The redshift range and the halo mass of SMGs also fits the cold accretion picture predicted by the simulations: they are commonly observed at $z \sim 2.5$ \citep{Chapman05}, and the clustering of SMGs indicate that they inhibit massive halos surpassing $M_{\rm shock}$ \citep[$M_{\rm DM} \sim 9 \times 10^{12}$ \msun;][]{Hickox12}. These distinctive attributes suggest that the halos of SMGs might encompass filaments of cool gas that permeate through the hot gas halo at the virial temperature. In the absence of such streams, the radiative cooling of the diffuse hot halo would transpire over a timescale that is considerably protracted, rendering it insufficient to sustain the observed high SFR. 

Despite their crucial role, definitive observational evidence is required to confirm their presence. This need grows stronger with recent simulation work showing cool streams become destabilized as they flow into hot halos due to Kelvin-Helmholtz Instabilities \citep{Mandelker19}. In addition, feedback processes occuring within the galaxy can disrupt the flows from the IGM \citep{Nelson15}. To verify their presence, a comprehensive examination into the CGM of these galaxies is necessary, and this investigation is typically conducted using two primary observational techniques. The first approach involves a pairing between emission selected foreground galaxies with bright Quasi-Stellar Objects (Quasars/QSOs) in the background as probes to study the CGM along their sightlines. This method leverages the absorption signatures imprinted by the intervening gas within the CGM onto the quasar spectrum, facilitating insights into the properties of the CGM. This methodology has been used for a variety of massive systems at $z \gtrsim 2$, such as QSOs \citep[e.g.,][]{Hennawi06,Prochaska13}, and SMGs \citep[e.g.,][]{Fu16,Fu21}. These close projections on the sky are rare, and we can only gather a one-dimensional view on the CGM. 

The second approach aims to detect and analyze the presence of extended Lyman-alpha (\Lya) emission within the CGM of these galaxies. A cold stream infalling towards a galaxy is expected to produce \Lya\ photons through gravitational heating \citep{Dijkstra09,Goerdt10,Rosdahl12}. In addition, background radiation \citep[i.e., the Cosmic UV Background; ][]{Gould96,Haardt12} can ionize the gas contained within the stream \citep{Faucher-Giguere10}, leading to \Lya\ recombination radiation from recaptured electrons. Consequently, a detection of \Lya\ filaments in surface brightness maps can serve as an indicator for cool streams. However, deep observations are required to detect the low surface brightness (SB) that is expected \citep{Rosdahl12,Faucher-Giguere10}. 

The diffuse emission can be enhanced by the presence of a QSO from photoionization and the resulting recombination radiation, increasing the surface brightness by a few orders of magnitude \citep{Cantalupo05,Cantalupo12,Kollmeier10,Hennawi13}. Furthermore, the resonant nature of the \Lya\ line can introduce scattering off of neutral hydrogen in the surrounding halo, producing a glow that can contribute to the observed \Lya\ luminosities measured in these halos \citep{Hennawi13,Gronke17,Byrohl22,Cen13}. Narrowband imaging campaigns targeting QSO systems at $z \sim 2 - 3$ began to find luminous \Lya\ halos with surface brightness $\gtrsim 10^{-17}$ \surf\ (\sbunit) \citep{Cantalupo14,Hennawi15,Cai17}. With the advent of integral field spectrographs (IFS) on 8-10 meter class telescopes, extended \Lya\ emission is now be regularly detected. Specifically, \Lya\ halos are now commonly found within QSO systems at $z \gtrsim 3$ \citep{Borisova16, Arrigoni-Battaia19,Fossati21} with the Multi Unit Spectroscopic Explorer \citep[MUSE;][]{Bacon10} and at $z \sim 2$ \citep{Cai19,OSullivan20a,Lau22,Vayner23} using the Keck Cosmic Web Imager \citep[KCWI;][]{Morrissey18}. MUSE has also detected \Lya\ halos in QSO systems out to $z = 6.6$ \citep{Farina19}. 

A selection of these observations have detected filamentary \Lya\ halos \citep{Cantalupo14,Hennawi15,Martin19}, suggesting the presence of cool streams. However, the presence of the QSO complicates the exact nature of those filaments. Outflows can become photoionized by the QSO and contribute to the observed \Lya\ emission \citep{Veilleux20}. Additionally, the expected surface brightness from recombination due to photoionization outperforms that of gravitational heating, so it is challenging to distinguish the two mechanisms.  We can counteract these challenges by focusing on systems devoid of AGN activity, such as the study performed by \cite{Daddi21}. The authors targeted a massive galaxy group with KCWI at $z \approx 2.9$ with no on-going AGN activity and found three extended filaments connecting at the center of the group; they conclude that the primary energy source for the observed \Lya\ halo is gravitational energy from infalling gas - a signature for cold accretion.

Observing campaigns targeting overdense "protocluster" regions have detected \Lya\ nebulae with physical sizes $\sim 100$\,kpc and total integrated \Lya\ luminosities ($L_{\rm Ly\alpha}$) $\gtrsim 10^{43}$\,erg s$^{-1}$, commonly referred to as "Lyman-alpha Blobs" (LABs) \citep{Steidel00, Matsuda04, Dey05, Nilsson06}. Gravitational energy from cold accretion has been proposed as a likely mechanism powering these bright nebulae. One such protocluster, the SSA22 field at $z\approx 3.1$, is home to several LABs \citep{Matsuda04,Matsuda11}. The famous LAB1 system \citep{Steidel00} is host to bright submillimeter sources, indicating elevated levels of star formation within dusty galaxies \citep{Chapman01, Geach14, Geach16,Umehata21}. While cold accretion may play a role, UV photons from the active galaxies may escape through unobscured sightlines and scatter back to the observer \citep{Geach14}. Additionally, the UV background is likely boosted by the ongoing star formation and AGN activity within the large protocluster environment. \cite{Umehata19} discovered filaments extending $\gtrsim 1$\,Mpc, likely illuminated by embedded sources. Other examples exist, such as the LAB within the luminous protocluster region at $z \sim 4$ hosting 10 dusty star-forming galaxies \citep{Oteo18}. In all of these examples, it is challenging to pinpoint a single dominant mechanism. We need to consider highly obscured targets that are relatively isolated (i.e., not in a clear protocluster environment). Specifically, SMGs have the potential to increase the contrast between the mechanisms powering \Lya\ emission.

The combination between an isolated source and the lack of a UV-continuum detected from ground surveys (due to high dust obscuration) poses a daunting challenge on to optical IFU surveys. Recent work by \cite{Lobos23a} found that extended \Lya\ nebulae around SMG-QSO composite systems are brighter than SMGs with no QSO companion. In fact, the latter do not show any extended \Lya\ emission above their surface brightness limit. Other examples include a gas-rich merger consisting of a QSO, Lyman-alpha emitters (LAEs), and a SMG was targeted at $z \approx 4.7$ by \cite{Drake20}. They detect a bright \Lya\ Halo around the QSO, but no associated \Lya\ nebula attributed to the SMG. These studies suggest that (1) detecting extended nebulae around SMGs require ultra-deep observations and (2) local ionization sources and \Lya\ emitters, such as the QSOs, play an important role in powering the extended nebula around QSOs.

Here we carry out deep integral-field spectroscopy of a rare SMG system, \system, where there are two QSOs in the background of the SMG. The \system\ system allows (1) a joint study of the \HI\ CGM in emission and in absorption and (2) a comparative study of the extended \HI\ \Lya\ nebulae around SMG and QSOs. The organization of this paper is as follows. We begin in \S\ref{sec:overview} by providing an overview of the \system\ system. We describe the KCWI observations, data reduction, and the procedure used to isolate extended \Lya\ signals in \S\ref{sec:Obs}. We present the resulting \Lya\ surface brightness maps and their kinematics for the SMG and the two background QSOs in \S\ref{sec:Results}. In \S\ref{sec:Compare}, we compare the \Lya\ emission lines against the \HI\ absorption lines towards the background QSOs and compare the three nebulae from this study with QSO extended \Lya\ nebulae in the literature. We conclude with a summary and give final remarks in Section \ref{sec:Summary}. 
We assume the $\Lambda$CDM cosmology with $\Omega_{\rm m}=0.3$, $\Omega_\Lambda=0.7$, and $h \equiv H_0/(100~{\rm km~s}^{-1}~{\rm Mpc}^{-1}) = 0.7$. Throughout we adopt the cgs units for surface brightness (\sbunit) and column densities (\pcm). 

\section{Overview of the \system\ System} \label{sec:overview}

\begin{figure*}[!htb]
    \centering
    \includegraphics[width=\textwidth]{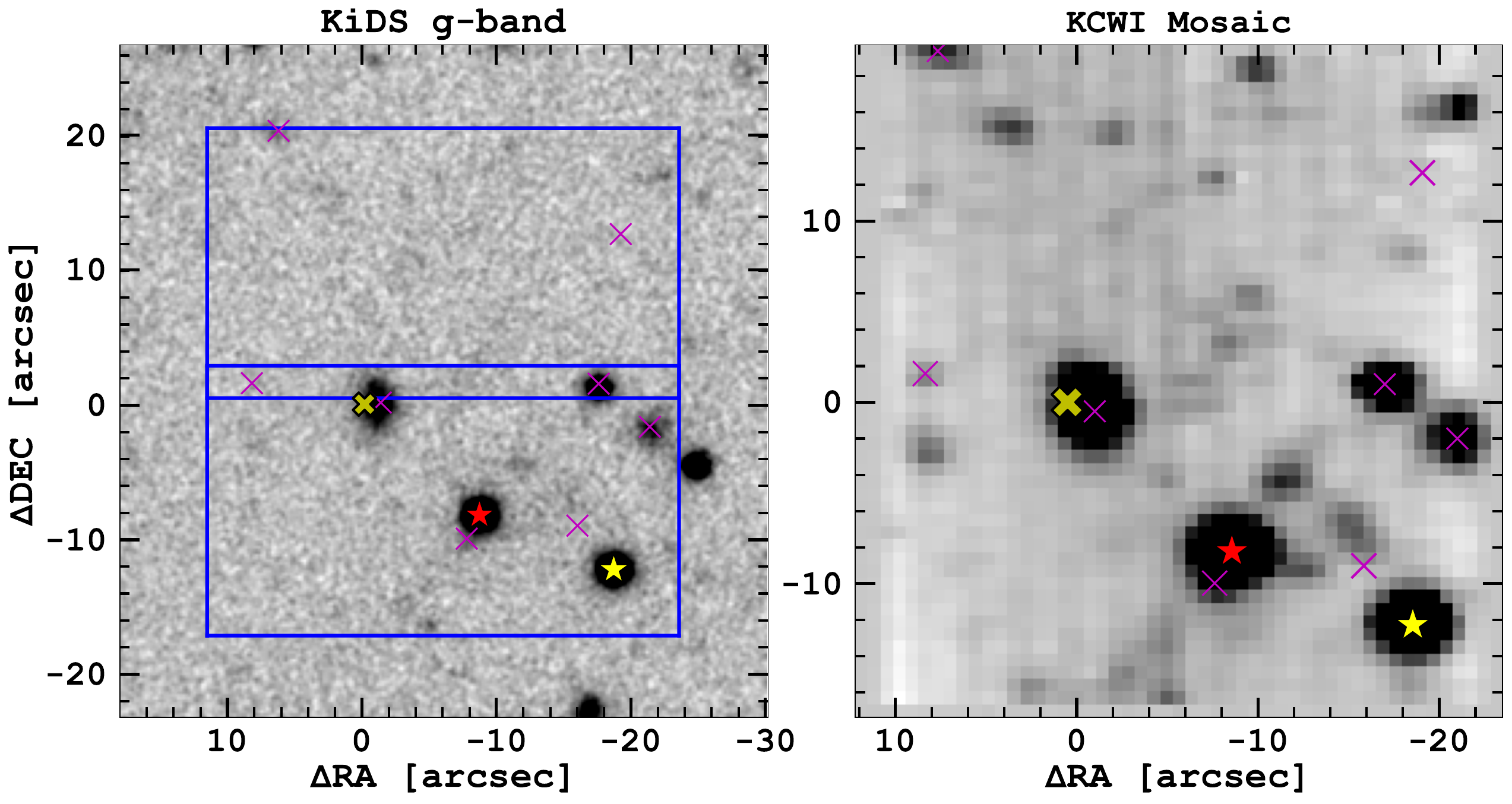}
    \caption{Overview of the \system\ field. On the left, we overlay the footprints of the two KCWI pointings (blue rectangles) on a deep $g$-band image (5$\sigma$ detection limit at 25.4\,mag) from the Kilo-Degree Survey \citep[KiDS;][]{de-Jong13}. On the right, we present the pseudo-$g$-band image from median averaging all wavelength channels of the coadded KCWI datacube. Although at a lower spatial resolution, the 2-hr-long KCWI image detects most of the sources present in the deep KiDS image. The positions of the SMG, \bgQSO\ and \fgQSO\ with a gold cross, a red star, and a yellow star, respectively. These same symbols will be used in subsequent figures. In addition to the three primary galaxies, we also mark the companion galaxies detected in \cothree\ (Comp a and b) and \Lya\ (LAE1 and LAE2), as well as four spectroscopically identified foreground interlopers (Obj 1-4).}
    \label{fig:finderchart}
\end{figure*}

This paper represents the fourth installment in a series of papers aimed at characterizing the CGM of SMGs with projected SMG-QSO pairs. Here we focus on the \system\ triplet system, which is a particularly rare alignment between a SMG (at $z_{\rm SMG} = 2.674$) and two background QSOs, both within a projected radius of 22\arcsec\ (see Figure\,\ref{fig:finderchart}). The triplet was initially identified in \cite{Fu16} by cross-matching large samples of spectroscopically confirmed QSOs with {\it Herschel}-selected 350\,\um\ peakers. Followup observations with ALMA in band-7 (345\,GHz/870\,\um) confirmed that the SMG exhibited an 870\,\um\ flux density of $S_{870} \approx$ 7.4\,mJy and was designated as \SMGfull\ (hereafter SMG; \citet{Fu17}). The improved FWHM resolution from ALMA band-7 imaging ($\sim$ 0\farcs5) provided accurate positioning for follow-up spectroscopy to acquire its redshift. In a focused study of the \system\ system, \citet{Fu21} (hereafter \citetalias{Fu21}), measured a redshift of $z = 2.674$ from \Ha\ with Gemini near-IR spectroscopy and \cothree\ with ALMA band-3 (100\,GHz/3\,mm) spectral imaging. They measured a total molecular gas mass of $M_{\rm mol} = 1.3 \times 10^{11}$\,\msun, representing one of the larger gas masses found among SMGs. Furthermore, the integrated IR Luminosity was $L_{\rm IR} \approx 1.2 \times 10^{13}$\,\lsun, which corresponds to a dust-obscured SFR of $1,200$\,\msunyr. Additionally, the ALMA 870\,\um\ and \cothree\ maps show no companions within $\lesssim 100$\,kpc. However, ALMA did detect three companion galaxies in \cothree: Comp a at $z = 2.6747$ and Comp b at $z = 2.6884, 2.6917$, but both reside $\gtrsim 150$\,kpc from the SMG. 

The SMG studied here is dominated by star formation with no major contribution from AGN activity. If we consider the spectral energy distribution (SED) of the SMG and compare with models for QSOs obscured by a dusty torus, as shown by \cite{Fu17}, the QSO+torus models underpredict the far-IR emission. Star formation sits as the dominate contributor to the bolometric luminosity. Although, SMGs are shown to harbor AGN activity through X-ray observations \citep{Alexander05, Wang13}, but no data is currently available.

The main text of the paper focuses on the three primary galaxies in the field (SMG, \bgQSO, and \fgQSO). For a complete inventory of spectroscopically identified objects in the field, refer to Figure\,\ref{fig:finderchart} for finder charts and Table\,\ref{tab:objects} for their coordinates and redshifts. In Appendix\,\ref{sec:low-z}, we show the spectra of the low-redshift interlopers (Objs 1-4), and Appendix\,\ref{sec:LAE2} gives a detailed discussion on the newly discovered \Lya\ emitter LAE2 and its implication to the \Lya\ absorber at $z_{\rm abs} \approx 2.69$ towards both QSOs. 

The two QSOs in the field have spectroscopic redshifts from the Sloan Digital Sky Survey (SDSS), and both lie in the background of the SMG. \bgQSOfull\ (hereafter \bgQSO, $g = 20.78$, $r = 20.38$, $L_{\rm bol} = 10^{46}$\,\ergps) at $z_{\rm QSO1} = 2.916$ has a projected separation of 11\farcs7 from the SMG, and \fgQSOfull\ (hereafter \fgQSO, $g = 20.71$, $r = 20.44$, $L_{\rm bol} = 10^{46}$\,\ergps) at $z_{\rm QSO2} = 2.749$ has a projected separation of 22\farcs1. At $z_{\rm SMG} = 2.674$, these positions correspond to impact parameters of 93.1 and 175.5\,kpc for \bgQSO\ and \fgQSO\ respectively, within the virial radius of a $10^{13}$\,\msun\ halo ($\sim$186\,kpc), providing a great opportunity to search for neutral hydrogen and metal ions in the CGM of the SMG.

Indeed, both \bgQSO\ and \fgQSO\ show strong \HI\ \Lya\ absorbers with velocity offsets of only $\delta v \approx -300$\,\kms\ from the SMG redshift. Towards \bgQSO, there is a sub-damped \Lya\ absorber (sub-DLA; $19 < \logNHI < 20.3$) with a total column density of $\logNHI \sim 20.2$. Towards \fgQSO, there is a Lyman limit system (LLS; $17.2 \leq \logNHI < 19$) with a total column density of $\logNHI \sim 18.6$. Although the sightlines are separated by 86\,kpc in transverse distance, the two absorbers show remarkable kinematic and metallicity coherence, suggesting that they belong to the same stream of gas. Their metallicity of [$\alpha$/H]\,$\approx -2$ places the absorbers near the 1$\sigma$ upper bound of the metallicity of the intergalactic medium (IGM), offering strong evidence of inflowing material from the cosmic web. Can we detect this metal-poor cool gas stream in emission? How does the CGM around SMG compare with the CGM around QSOs at a similar cosmic epoch? To address these questions, we have obtained deep integral-field spectroscopy of \system\ with the 10-meter Keck II telescope.

\begin{table}[]
    \centering
    \caption{Spectroscopically Identified Objects in the Field}
    \label{tab:objects}
    \begin{tabular}{lcccc}
        \hline
        \hline
        ID & R.A. (J2000) & Decl. (J2000) & $z_{\rm sys}$ & Comment \\
         & (h:m:s) & ($^\circ$:$\arcmin$:$\arcsec$) &  & \\
         \hline
         Primary Targets & & & & \\
         SMG & 09:13:39.55 & -01:06:56.4   & 2.674 & \cothree \\
         \fgQSO & 09:13:38.32 & -01:07:08.6  & 2.750 & \cothree \\
         \bgQSO & 09:13:38.98 & -01:07:04.5  & 2.916 & \Lya\ \\
         \hline
         High$-z$ Companions & & &  & \\
         Comp a & 09:13:38.28 & -01:06:43.8  & 2.675 & \cothree \\
         Comp b & 09:13:38.49 & -01:07:05.5   & 2.688 & \cothree \\
         & & & 2.692 & \cothree  \\
         LAE1 & 09:13:40.06 & -01:06:54.2  & 2.674 & \Lya \\
         LAE2 & 09:13:39.04 & -01:07:06.4  & 2.692 & \Lya \\
         \hline
         Low$-z$ Interlopers & & & & \\      
         \objone & 09:13:39.31 & -01:06:57.0   & 0.054 & H$\gamma$ \\
         \objtwo & 09:13:38.15 & -01:06:58.5   & 0.265 & [O\,II] \\
         \objthree & 09:13:38.41 & -01:06:55.5  & ...  & M5 star \\
         \objfour & 09:13:40.06 & -01:06:37.1   & 0.298 & [O\,II] \\
         \hline
         \hline
    \end{tabular}
    \tablecomments{Coordinates for the Primary Targets, Comp a, and Comp b come from ALMA imaging as reported in \citetalias{Fu21}; The remaining coordinates come from KCWI. We comment on the line used to measure $z_{\rm sys}$ for each target in the last column.}
\end{table}

\section{Observations and Data Analysis} \label{sec:Obs}

We carried out deep optical integral-field spectroscopy of \system\ with Keck II/KCWI on Jan 30, 2022 (UT). We used a single configuration throughout the night: the large image slicer, the blue medium grating tilted to a central wavelength of 4500\,\AA, and 2x2 binning on the CCD detector. With this configuration, we achieved a wavelength coverage between 4055\AA\ and 4940\,\AA\ (roughly $g$-band) with a dispersion of $\sim$0.5\,\AA/pix and a spectral resolution of $R \sim 2000$. The wavelength coverage includes \HI\ Ly$\alpha$ line between $2.33 < z < 3.06$ (enclosing the redshifts of the SMG and the two QSOs). Interloper galaxies emitting the \OII\ $\lambda\lambda$3727,30 doublet can also be identified between $0.088 < z < 0.325$. 

With the large image slicer at a position angle (PA) of $0^\circ$ (i.e., N is up), the field-of-view (FoV) is 33\farcs0 wide (24$\times$1.35\arcsec\ slits) and 20\farcs4 tall (the length of each slit) with an EW spatial resolution of 1\farcs35/slit and a NS spatial resolution of 0.294\arcsec/pix. Although all of the three primary targets (SMG, \bgQSO, \fgQSO) can fit within a single KCWI pointing, we added another pointing to detect the \Lya\ nebula to the north of the SMG (see the blue rectangles in the left panel of Figure\,\ref{fig:finderchart}). For the north pointing, we obtained $6 \times 20$ min (2\,hr) frames centered at a Right Ascension (R.A.) and Declination (Decl) of 09:13:38.98, $-$01:06:47.29. For the south pointing, we obtained $7 \times 20$ min (2.3 hr) frames centered at 09:13:38.98, $-$01:07:05.50. The north and south pointings are offset by $18\farcs0$, leaving a $2\farcs4$ overlapping region that accumulated a total integration time of 4.3 hours. To improve the spatial sampling in the EW direction, we dithered half a slit width ($0\farcs675$) between adjacent exposures. The combined FoV for our KCWI observation of \system\ is 33\arcsec\ wide by 38\arcsec\ tall. In the following, we describe the procedures of data reduction, mosaicing, continuum source subtraction, and optimal extraction of \Lya\ line emission.

\subsection{Datacube Construction and Mosaicing}

Data reduction was carried out with the IDL version of the KCWI data reduction pipeline\footnote{\url{https://github.com/Keck-DataReductionPipelines/KcwiDRP}} \citep{neill18}. We compared the reduced science frames using flat fields built with the internal lamp, the dome lamp, and the twilight, and adopted the calibrated frames using the dome flats because the sky-subtracted 2D spectra show the minimum level of systematic artifacts. Feige\,56 was used for flux calibration because it shows less absorption features in our wavelength range than Hiltner\,600 (the other standard star taken during the night). The end products of the pipeline are fully calibrated datacubes $(x,y,
\lambda)$ that are also corrected for differential atmospheric refraction, along with the associated variance cubes. 

To combine the individual datacubes, we first need to solve for the astrometry by matching the positions of the sources detected in the wavelength-collapsed KCWI pseudo-$g$-band image with their equatorial coordinates from the KiDs catalog. The north pointing contains few bright sources, so we aligned these frames based on a single point source in the overlapping region between the pointings (Obj\,3 - a M5 star). For the south pointing, additional objects such as the QSOs were used to improve the astrometric solution. 

To coadd the datacubes based on their astrometric solutions, we used a modified version of the Cosmic Web Imager Tools \citep[CWITools; ][]{OSullivan20}. The python package combines the datacubes by re-sampling the pixels onto a common grid using the {\it drizzle} algorithm \citep{Fruchter02}. We chose a final pixel size of $0\farcs7 \times 0\farcs7$ to match the typical seeing and the 1\farcs35 slit width; any pixel size smaller led to noticeable artifacts. To illustrate the coadding result, we produced a pseudo-$g$-band image from the coadded KCWI datacube by median-averaging all of the wavelength channels. As Figure\,\ref{fig:finderchart} shows, this broad-band image detects almost all of the sources detected in the deep $g$-band image from KiDS that reached a 5$\sigma$ detection limit of 25.4 magnitude, despite its lower resolution. 

To produce the coadded variance cube for S/N estimation, we first pass the variance cubes through the same {\it drizzle} procedure just like the flux cubes. To account for covariances introduced to the coadded datacube from resampling, we then use the CWITools covariance estimation function to compute the factor that should be multiplied to the resampled variance cube \citep[refer to Section 5.2 in ][for details]{OSullivan20a}. For our case, we found a multiplication factor of $\sim 2$.

\subsection{QSO and Continuum Subtraction} \label{sec:QSO_cont_sub}

In order to isolate the extended \Lya\ emission, we need to remove the bright continuum sources in the datacubes. We found that better results can be obtained by removing continuum sources in individual datacubes and then coadding the resulting cubes, as opposed to removing continuum sources from the coadded datacube. Note that the datacubes from individual exposures preserve the intial spatial sampling and have a pixel size of 1\farcs35$\times$0\farcs294. The coadding procedure is the same as outlined in the previous subsection and the coadded datacube has a pixel size of 0\farcs7$\times$0\farcs7.

Inspired by previous analytical methods from optical IFU studies of \Lya\ nebula \citep[][and references therein]{Borisova16,Arrigoni-Battaia19,Cai19,OSullivan20a,Lobos23a}, we first subtract the bright QSOs with empirical PSF models. For both PSF model construction and subtraction, we only focus on the subregion of $4\farcs1 \times 4\farcs4$ ($3 \times 15$ pixels) centered on each QSO, because the size is large enough to capture most of the flux from a point source. We build a high S/N PSF model for each wavelength channel by median-combine the adjacent 300 channels (150\,\AA), and subtract the PSF out by scaling it to the flux within the central $1\farcs4 \times 1\farcs5$ ($1 \times 5$ pixels) region of each QSO. There are a couple of caveats in this process: (1) the \bgQSO\ completely disappears in certain wavelength channels because of the DLAs, making QSO subtraction unnecessary in these channels; we also do not use these channels when building the PSF because they contain no PSF signal. (2) The wavelength channels near the systemic redshifts of the three primary galaxies should not be used to build the PSF model because they may be contaminated by the extended \Lya\ emission that we are after. Our analysis code is flexible enough to account for these caveats. 

Next, we subtract the remaining continuum sources with a running average method. It is necessary to separate each QSO subtraction from continuum subtraction because the QSOs exhibit sharp spectral features in both emission and absorption. The main difference is that we build the continuum model for the full spatial coverage of each datacube and subtract the continuum model without scaling. Similar to the QSO subtraction, we build a high S/N continuum model for each wavelength channel by median-combine the adjacent 300 channels, excluding the channels near the systemic redshifts of the primary galaxies. This process removed most of the continuum sources seen in Figure\,\ref{fig:finderchart}, but it did not work well for sources that show narrow emission lines or other sharp spectral features. So we mask out these high residual foreground sources in the maps that we will present.

\begin{figure*}[!htb]
    \centering
    \includegraphics[width=\textwidth]{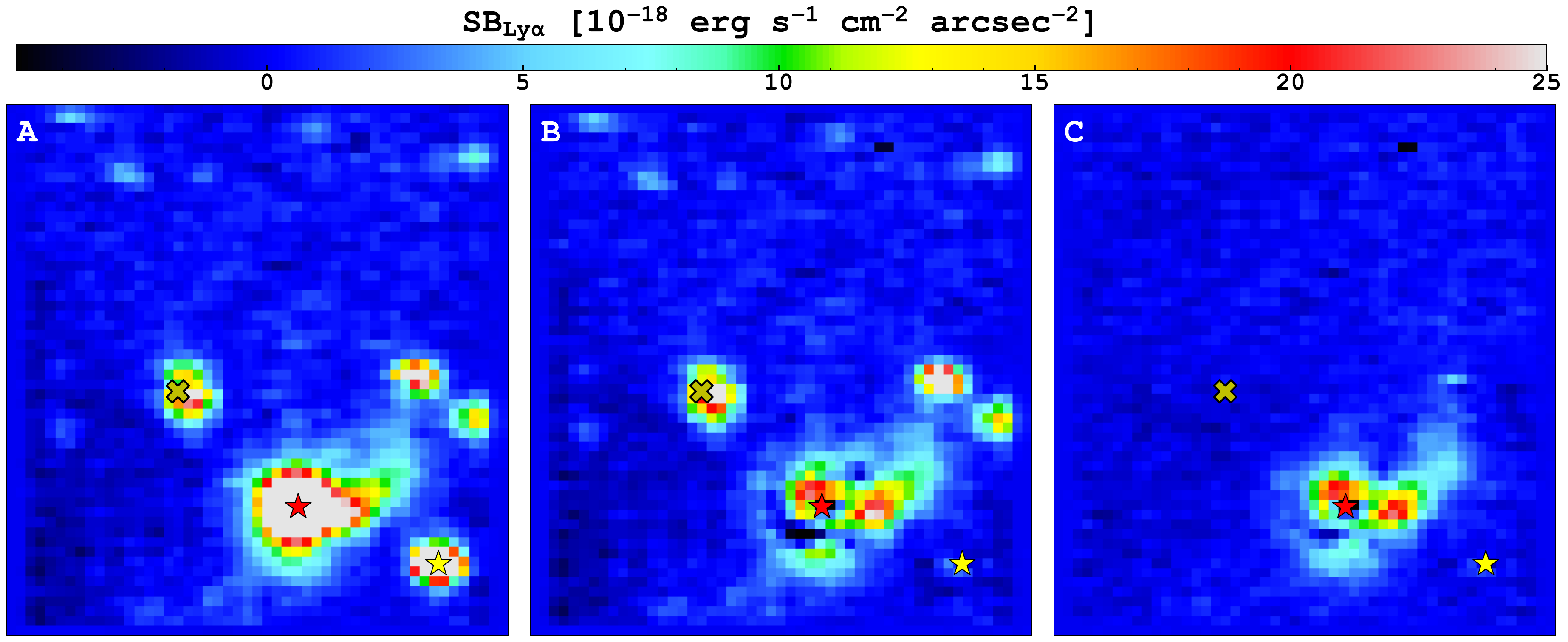}
    \caption{Demonstration of QSO- and continuum-subtraction results. Each panel displays a narrow-band image formed by combining wavelength channels within $\pm 1500$\,\kms\ relative to $z_{\rm QSO1}$. From left to right, we show the narrow-band image from the datacube before any source subtraction, that from the datacube after QSO subtraction, and that from the datacube after both QSO and continuum subtraction. Only the extended \Lya\ nebula remains after the subtractions.}
    \label{fig:cube_process}
\end{figure*}

Finally, once the QSOs and other continuum sources have been subtracted from each individual datacube, we combine the cubes with the same astrometry solutions and the same {\it drizzle} procedure in the previous subsection. To illustrate the performance of the subtraction procedure, we generated narrow-band images optimized for extended \Lya\ nebula around \bgQSO by summing up the wavelength channels within $\pm 1500$\,\kms\ relative to $z_{\rm QSO1} = 2.916$. Figure\,\ref{fig:cube_process} shows the narrow-band images from the coadded datacubes before and after each subtraction stage. It is clear that both the QSO and continuum sources are adequately removed from the final coadded datacube. More importantly, the process unveiled a giant \Lya\ nebula around \bgQSO, which we will discuss in more detail along with the nebulosity around the SMG and the \fgQSO\ in the next subsections.  

\subsection{Optimal Extraction of \Lya\ Emission} \label{sec:optextract}

Because extended \Lya\ line emission display complex kinematics with a variety of velocities and velocity widths, simply combining wavelength channels within a fixed velocity range around a systemic redshift (like we did in Figure\,\ref{fig:cube_process}) does not provide the optimal result. Optimization is especially important for recovering low S/N signals around high redshift galaxies. The ``optimal extraction'' approach solves this problem by finding an optimal extraction window at each spatial location, and various techniques have been applied on datacubes produced by both radio interferometers \citep[e.g.,][]{Wong22}\footnote{\url{https://github.com/tonywong94/maskmoment}} and optical integral field spectrographs \citep[e.g.,][]{Arrigoni-Battaia19,Cai19}. These extraction windows together is a 3-dimensional mask generated from a smoothed S/N cube. We begin by constructing a 3D Gaussian kernel to smooth the the QSO- and continuum-subtracted coadded datacube. For the spatial dimensions, we use a 2D Gaussian Kernel with a FWHM of 1.5\arcsec, comparable to the spatial resolution. For the spectral dimension, we used a Gaussian with a FWHM matching the spectral resolution (2.25\AA). The kernel size is $7 \times 7 \times 5$ pixels ($4\farcs9 \times 4\farcs9 \times 2.5$\,\AA). We convolve the 3D kernel with both the flux cube and the variance cube and divide the two to build a smoothed S/N cube. All pixels with a S/N less than S/N$_{\rm min}$ are masked out. We then employ the Python module \texttt{cc3d}\footnote{\url{https://pypi.org/project/connected-components-3d/}} to discern and categorize all clusters of pixels exhibiting three-dimensional spatial connectivity. We then exclude clusters with less than $N_{\rm min}$ pixels to minimize the number of spurious ``floating islands''. The surviving clusters form a 3D mask that can be used to extract line emission from the {\it unsmoothed} datacube to generate the ``optimally extracted'' \Lya\ surface brightness maps and its kinematics maps. 

However, the parameters described above ($N_{\rm min}$ and S/N$_{\rm min}$) need to be chosen to optimize the 3D mask. For example, a too ``inclusive'' 3D mask would not only reduce the S/N of the extracted maps but also overestimate the size of the nebula. We thus devised an iterative procedure to find the parameters that optimize the 3D mask. For each primary target, we start the iteration by guessing the initial values of $N_{\rm min}$ and S/N$_{\rm min}$. Once the initial 3D mask is generated, we apply it to the {\it unsmoothed} flux cube and variance cube and generate a S/N map for the \Lya\ emission. We then compare the S/N map with the projected boundaries of the 3D mask. Because only pixels with S/N\,$\gtrsim 2$ should be considered as detection, if the projected boundaries of the 3D mask include too many pixels with S/N\,$< 2$, a more ``exclusive'' mask should considered. In the next iteration, the parameters $N_{\rm min}$ and S/N$_{\rm min}$ are modified accordingly and a new 3D datacube produces a new S/N map and the 2$\sigma$ contour is again compared with the projected boundaries of the mask. After a few iterations, the mask is considered ``optimal'' when a good agreement is found between the two. Through this process, we found that the optimal parameters are $N_{\rm min} = 800$ and S/N$_{\rm min} = 1.2$.

\section{Results} \label{sec:Results}

Both SMGs and QSOs represent the most massive halos ($10^{12.5-13}$\,\msun) in the early universe, as indicated by their strong spatial clustering signals \citep[e.g.,][]{Hickox12,DaAngela07,Trainor12,White12}. By design, our KCWI data covers the redshifted \Lya\ line around all three primary galaxies, making it possible to directly compare their observed properties at the same depth and with the same instrument. We present the \Lya\ results from three different angles. First, we build \Lya\ channel maps to illustrate the emission detected in the ``raw'' data. Then, we present optimally extracted \Lya\ moment maps to showcase each nebula in more detail. Finally, we construct azimuthally averaged radial profiles of the \Lya\ surface brightness to quantify the similarities and differences among the three galaxies. 

\subsection{\Lya\ Channel Maps} \label{sec:ChannelMaps}

\begin{figure*}[t]
    \centering
    \includegraphics[width=1.\textwidth]{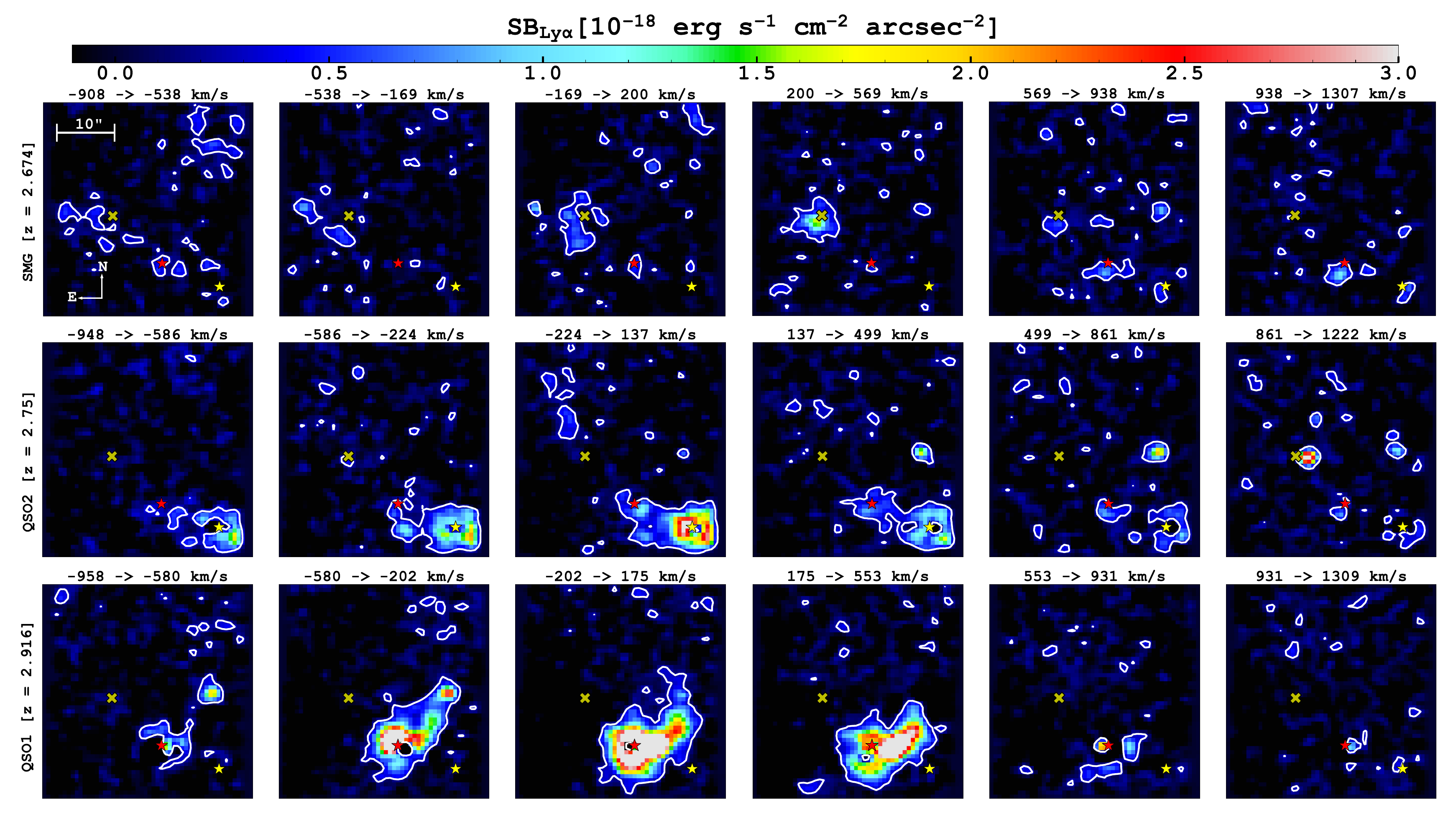}
    \caption{\Lya\ channel maps. The top panel displays the SMG, and each map has a velocity width of $\approx 150$\,\kms. The middle and bottom panel show the channel maps for \fgQSO\ and \bgQSO\ respectively with each velocity width increased to $\approx 650$\,\kms. The white contours corresponds to $\rm S/N = 2$. The position of the SMG, \bgQSO, and \fgQSO\ are represented by the gold cross, red, and yellow stars respectively. }
    \label{fig:channel_maps}
\end{figure*}

We begin our search for extended \Lya\ emission by producing \Lya\ channel maps from the QSO- and continuum-subtracted datacube. For each target, we produce a series of narrow-band images spanning a velocity range of $\sim 1300$\,\kms\ relative to the systemic redshift, each covering a velocity range of $\sim$370\,\kms\ (Figure\,\ref{fig:channel_maps}). We estimate the 1$\sigma$ noise of each image by computing the root mean square (RMS) from near empty regions, and overlaid the 2$\sigma$ contours on each image. Extended \Lya\ emission is detected around all three galaxies, and each nebula cover quite a large velocity space. From these channel maps, it is already clear that the QSOs exhibit significantly brighter extended \Lya\ emission than the SMG. 

Around the SMG, there is a bright \Lya\ nebula near the SMG position and $\delta v = 200$\,\kms, and there is also patchy filamentary emission emanating from the SMG along the diagonal directions in the other channel maps. Next up in redshift, around \fgQSO, the channel maps reveal a bright circular \Lya\ blob centered on the QSO and the emission clearly extends diagonally towards \bgQSO. This extended emission is likely connected to the DLA at $z_{\rm abs} = 2.75$ towards \bgQSO. Lastly, around \bgQSO, we detect the brightest \Lya\ nebula among the three galaxies. Similar to the other two galaxies, we also observe a bright central \Lya\ blob with a filamentary extension (this time towards the NW direction). 

\subsection{Optimally Extracted \Lya\ Moment Maps} \label{sec:LyaMomMaps}

\begin{figure*}[!ht]
    \includegraphics[width=\textwidth]{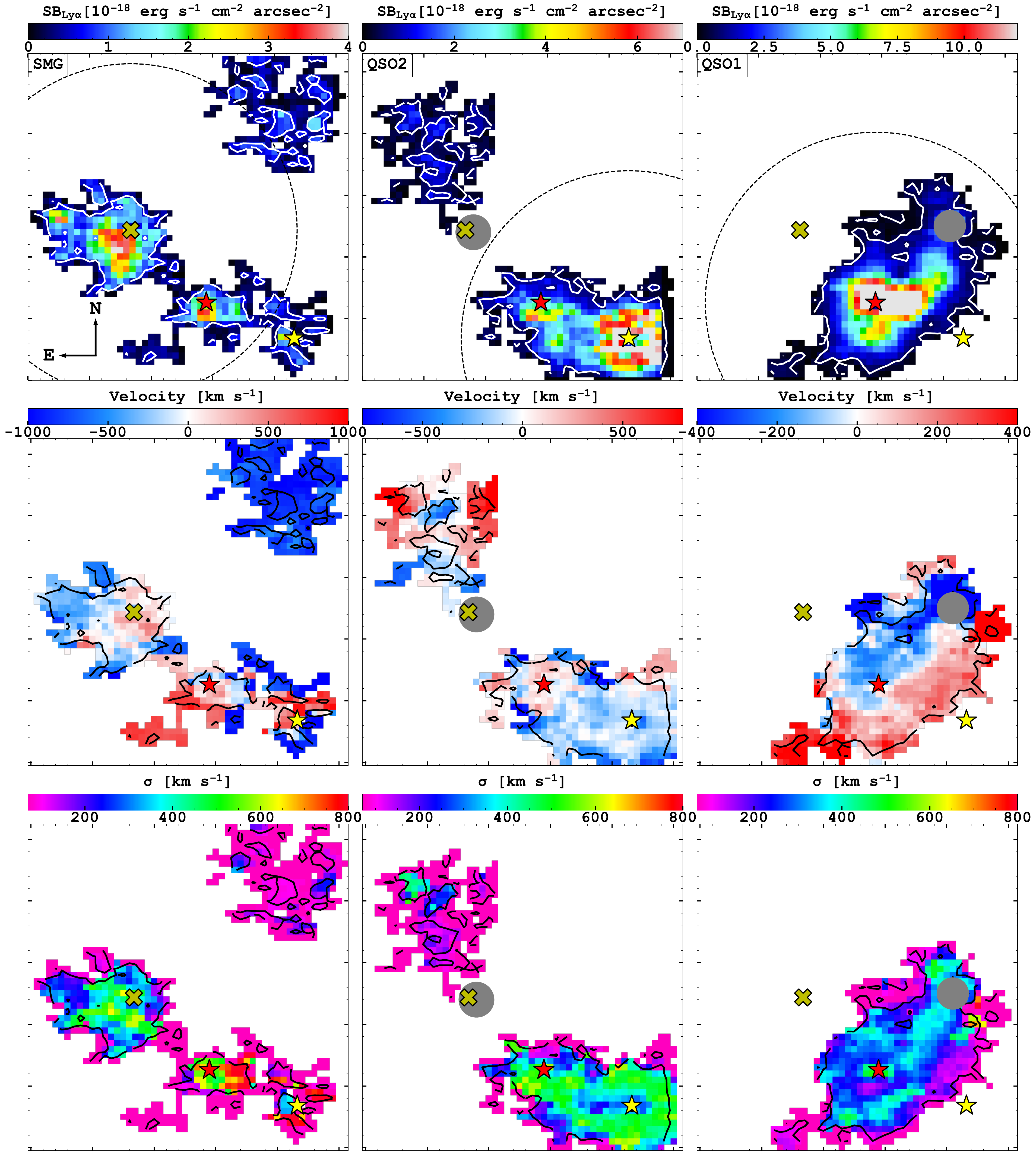}
    \caption{Moment maps of extended \Lya\ emission around the three primary galaxies: SMG, \fgQSO, and \bgQSO\ from left to right, and the \Lya\ surface brightness maps, velocity maps, and velocity dispersion map from top to bottom. For all images, we overlay the $\rm S/N = 2$ contours. To indicate the scale, a dashed circle with a radius of $150$\,kpc is drawn around each host galaxy. The gray circles are masked out areas due to foreground objects.}
    \label{fig:moment_maps}
\end{figure*}

\begin{table}[]
    \centering
    \caption{Properties of Extended \Lya\ Nebulae}
    \label{tab:lya_results}
    \begin{tabular}{lcccr}
        \hline
        Parameter & SMG & \fgQSO\ & \bgQSO\ & Unit  \\
        \hline
        \hline
        Luminosity & $1.1 \pm 0.02$ & $3.4 \pm 0.07$ & $7.1 \pm 0.04$ & $10^{43}$\,\ergps \\
        Area       & 67 (4200) & 113 (7020) & 159 (9600) & arcsec$^2$ (kpc$^2$) \\
        Extent     & 31 (250) & 38 (300) & 26 (200)  & arcsec (kpc) \\
        Peak SB    & 4.5 & 12 & 27 & $10^{-18}$ \surf \\
        Mean SB    & 1.0 & 1.9 & 3.6  & $10^{-18}$ \surf \\
        $\langle V \rangle$ & -250 & -51 & 2.4 & \kms \\
        $\langle \sigma \rangle$ & 180 & 240 & 170 & \kms \\
        Profile Index & $-0.67\pm0.22$ & $-1.6\pm0.3$ & $-1.9\pm0.3$ & --- \\
        \hline
    \end{tabular}
    \tablecomments{All measured within the $2\sigma$ contours of the optimally extracted surface brightness maps. }
\end{table}

To characterize the extended \Lya\ emission at the highest possible S/N, we utilize the 3D optimal extraction masks described in \S\,\ref{sec:optextract}. While the channel maps provide crude information on the gas morphology and kinematics, the optimally extracted emission line maps in Figure\,\ref{fig:moment_maps} provide the full morphology of each nebula, the total \Lya\ Luminosity ($L_{\rm Ly\alpha}$), the total physical size, and detailed kinematics. As usual, the moment 0, 1, and 2 maps are the \Lya\ surface brightness maps, the flux-weighted \Lya\ radial velocity, and the velocity dispersion maps. All maps are generated using the optimized 3D mask and the {\it unsmoothed}, QSO- and continuum-subtracted, coadded flux cube. Corresponding noise maps in surface brightness are from the coadded variance cube. The reader can directly compare these maps with the channel maps in Figure\,\ref{fig:channel_maps} to evaluate the likelihood of spurious features. The three galaxies display a variety of \Lya\ morphology and kinematics as we already started noticing in the channel maps. The area around \bgQSO\ shows nebulae at a number of redshifts, we rule out the possibility of mis-identification in Appendix\,\ref{sec:QSO1_south}. In the following, we describe the nebula around each galaxy in the order of increasing redshift. Table\,\ref{tab:lya_results} summarizes the properties of the \Lya\ nebulae measured within the $2\sigma$ detection contours.

\subsubsection{SMG at $z = 2.674$}

With a $2\sigma$ detection limit of $6.7 \times 10^{-19}$\,\surf, we detect a clear extended \Lya\ nebula surrounding the SMG. Similar to the \Lya\ channel maps in \S \ref{sec:ChannelMaps}, the nebula possesses the brightest SB near the position of the SMG reaching a peak value of $\approx 4 \times 10^{-18}$\,\surf. Shifting $\sim 10$\arcsec\ to the northeast of the SMG, we detect the presence of LAE1 by identifying a local increase in SB up to $\approx 3 \times 10^{-18}$\,\surf. We also detect clumpy \Lya\ emission nearly $16$\arcsec\ to the northwest of the SMG reaching a peak SB of $\sim 1.5 \times 10^{-18}$\,\surf. From the SB map alone, it is unclear whether this clumpy emission is associated with the SMG or other independent sources. 

As we shift towards the southwest direction, we identify a large filamentary structure stretching over $180$\,kpc from the SMG intercepting the sightlines of both QSOs. The structure is not a single solid filament, but a collection of large regions of clumpy \Lya\ emission that takes the shape of a single filament. The mean SB along the filament is $\sim 1.2\times 10^{-18}$\,\surf. The presence of this large structure and its position in the CGM of the SMG suggests a connection with the absorption systems detected previously for both QSOs.

The total nebula area captured within our $2\sigma$ contour is $67$\,arcsec$^2$ ($4200$\,kpc$^2$). We measure the total extent of the nebula to be $31$\arcsec\ (250 kpc). The mean SB throughout the entire nebula is $1.0 \times 10^{-18}$\,\surf, and the total integrated $L_{\rm Ly\alpha}$ is $1.1 \pm 0.02 \times 10^{43}$\,erg s$^{-1}$. 

Moving onto the kinematics of the gas, we detect a velocity gradient within the brighter region of the nebula. The \Lya\ line at the position of the SMG resides at $\delta v \approx +300$\,\kms. As we shift to the East, the gas gradually blueshifts to $\delta v \approx -300$\,\kms. The large collection of \Lya\ clumps to the northwest of the SMG all exhibit blueshifts $\lesssim -700$\,\kms, suggesting that they are outside the CGM of the SMG. The large filamentary structure possesses both redshifted and blueshifted components with no velocities exceeding an absolute value of $700$\,\kms. From the moment 2 map, we find that all regions captured within the $2\sigma$ contour possess line widths $\sigma \gtrsim 150$\,\kms. Regions that show larger line widths ($\sigma \geq 700$\,\kms) are the result of blended blueshifted and redshifted components (this will be true for both QSO nebulae). Overall, we measure an average velocity offset of $-250$\,\kms\ and an average line width of $180$\,\kms\ for the entire nebula.

\subsubsection{\fgQSO\ at $z = 2.75$}

Continuing onto \fgQSO, we detect a \Lya\ nebula extended throughout its CGM at a $2\sigma$ detection limit of $6.3 \times 10^{-19}$\,\surf. Similar to the SMG, we find the brightest \Lya\ SB localized near the position of the galaxy, which reaches a peak value of $12.0 \times 10^{-18}$\,\surf. As we shift away from the QSO, we see a large filamentary structure extending nearly $\sim 150$\,kpc to the northeast intercepting the sightline of \bgQSO. The mean SB along this filament decreases down to $\approx 2.0 \times 10^{-18}$\,\surf. Near the position of \bgQSO, we measure an increase in SB values approaching $\approx 5\times 10^{-18}$\,\surf. The filament disappears from view only a few arcsecs ($\sim 20$\,kpc) to the northeast of \bgQSO's sightline. 

Looking even further to the northeast, we detect a pocket of clumpy \Lya\ emission with a mean SB of $\approx 1 \times 10^{-18}$\,\surf. Interestingly, the clumps take the shape of a narrower filament in a similar fashion to the structure observed in the CGM of the SMG. Intriguingly, We notice that this extended structure is positioned in such as way to form an arc with the filament seen in the southwest. This suggests that this could be one large extended filamentary structure ($\gtrsim 200$\,kpc) within the CGM of \fgQSO\ casting a shadow onto \bgQSO's sightline (the DLA at $z=2.75$). The obvious gap in the filament may be due to the presence of Obj 1, which we masked out due its strong H$\gamma$ line that was redshifted into the wavelength window of \fgQSO's \Lya\ line. 

If we assume one large structure, the total extent of \fgQSO's nebula is nearly $38$\arcsec\ ($300$\,kpc), noticeably larger than the SMG nebula. Furthermore, the total detected nebula area is $113$\,arcsec$^2$ ($7020$\,kpc$^2$). The mean \Lya\ SB is $1.9\times 10^{-18}$\,\surf, and the total integrated $L_{\rm Ly\alpha}$ is $3.4 \pm 0.07 \times 10^{43}$\,erg s$^{-1}$. 

The kinematics seen in \fgQSO's nebula also show a large distribution of radial velocities. Near the position of the galaxy, we find velocities within a range between $-600 < \delta v < 0$\,\kms. As we shift towards the position of \bgQSO, we measure velocities $0 < \delta v < +300$\,\kms. The extended structure to the northeast does not possess a specific $\delta v$. Rather, we measure a range of $-500 < \delta v < +600$\,\kms. The mean velocity offset of the full nebula is $\approx -50$\,\kms. Using the moment 2 map, we find that all detected \Lya\ emission exhibits line widths larger than $\sigma \gtrsim 180$\,\kms. The mean line width across the full nebula is $240$\,\kms. Similar to the SMG's nebula, we do not detect any clear velocity structure within \fgQSO's nebula.

\subsubsection{\bgQSO\ at $z = 2.916$}

Concluding with \bgQSO, we detect the brightest \Lya\ nebula out of the three primary galaxies at a $2\sigma$ detection limit of $6.0 \times 10^{-19}$\,\surf. The peak \Lya\ SB is $\approx 27 \times 10^{-18}$ \surf, which is $\approx 2.3 \times$ higher than \fgQSO\ and $6\times$ higher than the SMG. The mean SB of the entire nebula is $\sim 3.6 \times 10^{-18}$ \surf. Similar to \fgQSO, a rather large structure can be seen over $100$\,kpc to the northwest of the QSO with a SB $\gtrsim 4 \times 10^{-18}$\,\surf. Given the similarity between the structures seen in the CGM of both QSO's, it would be remarkable if both structures are DLA systems. However, we are unable to speculate since there is no background target intercepted by the structure in \bgQSO's CGM. 

The nebula found within the \bgQSO's CGM resembles previously observed QSO \Lya\ nebulae, particularly those classified as enormous \Lya\ nebula (ELAN) \citep{Cantalupo14,Hennawi15,Cai17,Cai18} that have \Lya\ surface brightness $\gtrsim 1\times10^{-17}$\,\surf\ spanning 100s of kpc with a total integrated $L_{\rm Ly\alpha} \geq 10^{44}$\,erg s$^{-1}$. However, we measure $L_{\rm Ly\alpha} \approx 7 \times 10^{43}$ \ergps\ for the entire \Lya\ halo. The total nebula area is $\approx 159\,\rm arcsec^2$ ($9600\,\rm kpc^2$). 

Unlike the previous galaxies, we find notable kinematic features present within \bgQSO's nebula. Two distinct velocity components, each with $|\delta v| \approx 300$\,\kms, are detected within the nebula. If we compute the mean velocity offset for the nebula, we find that these velocity components average out to $\sim 0$\,\kms. \bgQSO's position coincides with the boundary separating these velocity regions, spanning the majority of the observed extended nebular structure. Furthermore, a systematic increase of approximately $50-100$\,\kms in the absolute velocity value is observed as one moves away from the galaxy's position. From the moment 2 map, we find that the mean \Lya\ line width in the nebula is $\sigma = 170$\,\kms. Regions with $\sigma \gtrsim 450$\,\kms\ correspond to the boundary between both velocity components. However, we find line widths approaching $\sim 400$\,\kms\ $\approx 5$\arcsec\ to the south of \bgQSO, which may not be influenced by line-blending. In Appendix \ref{sec:QSO1_vel}, we discuss the remarkable similarities between \bgQSO's kinematic features with those detected in recent observations of the MAMMOTH-1 \Lya\ nebula \citep{Zhang23}. Overall, the kinematics of \bgQSO\ offer a greater insight into the dynamics of the cool gas when compared with the SMG and \fgQSO.

\subsection{Azimuthally Averaged Radial Profiles} \label{sec:sbprofiles}

\begin{figure}[!ht]
    \hspace*{-0.5cm}       
    \includegraphics[width=0.5\textwidth]{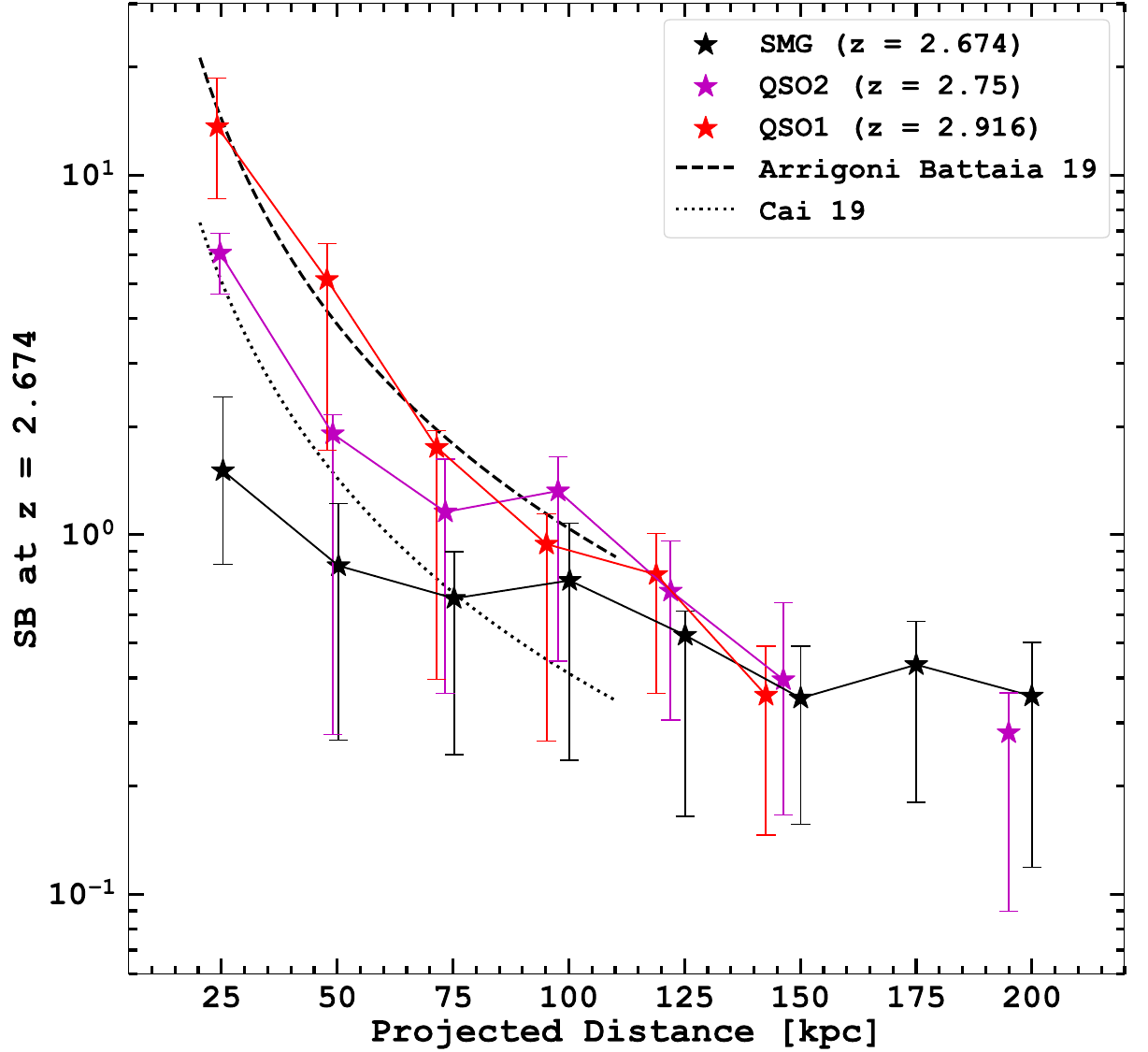}
    \caption{Azimuthally averaged \Lya\ surface brightness profiles. The nebulae around the SMG, the \fgQSO, and the \bgQSO\ are plotted in black, purple, and red, respectively. The error bars correspond to the 25th and 75th percentile of the pixels within each annulus. For comparison, we also include the average profiles of previous QSO surveys at $z \sim 3$ \citep{Arrigoni-Battaia19} and $z \sim 2$ \citep{Cai19} as dashed and dotted line, respectively. All surface brightnesses are scaled to the SMG redshift at $z_{\rm SMG} = 2.674$.}
    \label{fig:SB_profile}
\end{figure}

We now compare the radial profiles of the three nebulae by constructing azimuthally averaged surface brightness profiles from the moment 0 maps. Without information on the inclination angle of the systems relative to the observer, we simply generate evenly spaced circular annuli around each primary galaxy and average the \Lya\ surface brightness from the pixels in each annulus. The annuli begin at $\sim$25\,kpc and end at $\sim$200\,kpc, to encompass the expected virial radii of $10^{13}$\,\msun\ halos at $z \sim 2.6-2.9$. Because of the clumpy morphologies of the nebulae, not all annuli have valid measurements. Because of the cosmological surface brightness dimming ($S(z) \propto (1+z)^{-4}$), the radial surface brightness profiles of the QSOs are scaled to the SMG redshift by multiplying $(1+z_{\rm QSO})^4/(1+2.674)^4$. 

Figure\,\ref{fig:SB_profile} shows the resulting surface brightness profiles. In the inner 100\,kpc or so, the QSOs clearly show an enhanced surface brightness profile; but in the outer region, all profiles converge to a plateau around $4\times10^{-19}$\,\surf. Simple power-law model fits to the radial profiles shows that the power-law index steepens from $-0.67\pm0.22$ around the SMG, $-1.6\pm0.3$ around \fgQSO, to $-1.9\pm0.3$ around \bgQSO. As shown in the figure, the radial profiles of two QSO nebulae are comparable to those measured previously around other QSOs \citep{Arrigoni-Battaia19,Cai19}, but the profile of the SMG nebula is significantly shallower. 

\section{Comparisons} \label{sec:Compare}

In this section, we first compare the extended \Lya\ emission line detected near each QSO with the \HI\ absorption profiles present in the QSO spectra to connect the ionized hydrogen with the neutral hydrogen in the gas stream around the SMG, and then compare the general properties of the \Lya\ nebulae from this study to those of \Lya\ nebulae from the literature. 

\subsection{\HI\ Emission vs. Absorption} \label{sec:Emission_Absorption}

\begin{figure*}[!htb]
    \centering
    \includegraphics[width=\textwidth]{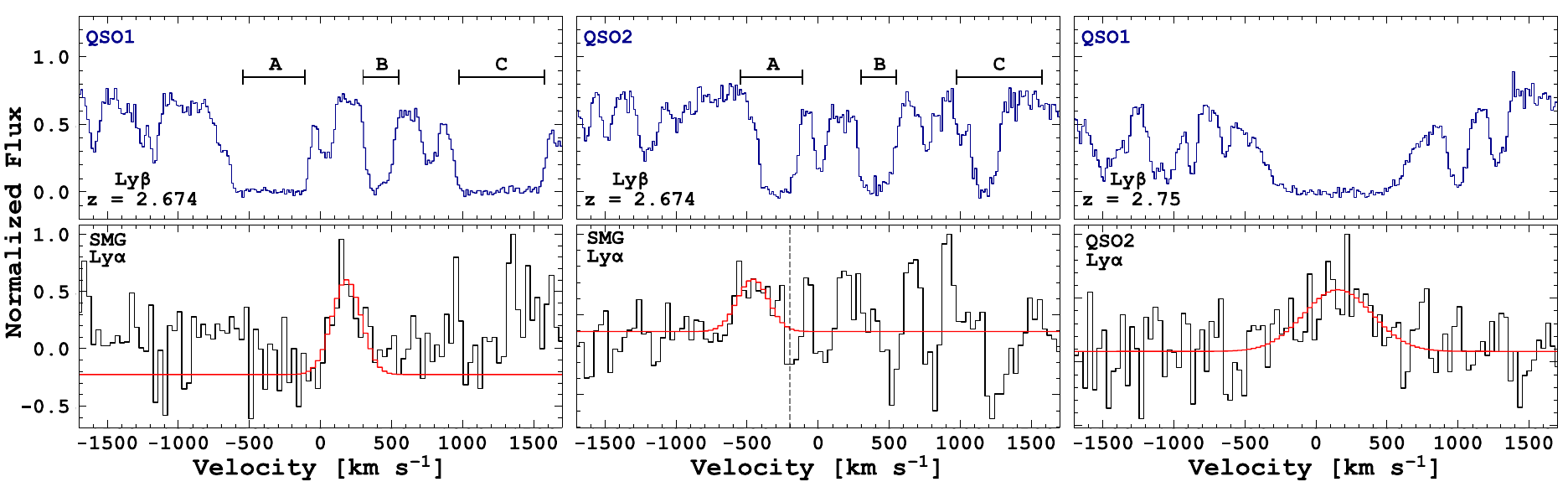}
    \caption{\HI\ \Lyb\ absorption towards QSO sightlines ({\it top}) vs. extended \HI\ \Lya\ emission near the sightlines ({\it bottom}). The continuum-normalized QSO spectra on the top are from VLT/X-shooter, while the \Lya\ emission-line spectra at the bottom are extracted from our QSO and continuum-subtracted KCWI datacube. All velocities are with respect to the redshift denoted in the top right of each column. For each \Lya\ emission line spectrum, we overlay the best-fit Gaussian profile in red. {\it Left:} \HI\ absorption/emission towards/near \bgQSO\ around the SMG redshift. {\it Middle:} \HI\ absorption/emission towards/near \fgQSO\ around the SMG redshift. {\it Right:} \HI\ absorption/emission towards/near \bgQSO\ around the \fgQSO\ redshift. The dashed line seen in the middle panel corresponds to the velocity beyond which QSO subtraction is important; this explains the large residuals at these velocities. The three cases display a range of velocity offsets between the main absorption clouds and the adjacent emission-line nebulae.}
    \label{fig:emission_absorption}
\end{figure*}

While QSO absorption line spectroscopy is an extremely sensitive tool to detect intervening clouds and their chemical composition, the spatial extent and distribution of the clouds are difficult to quantify even with multiple QSO sightlines. A complete picture of the cool gas in the CGM requires deep integral-field spectroscopy that is sensitive enough to detect diffuse low surface-brightness line emission. Now with the KCWI data, we can begin to piece the pictures together. 

For the \system\ system, there are two main filaments of cool gas for which we can compare the gas kinematics in absorption and in emission: (1) the stream around the SMG that is responsible for the $z_{\rm abs} \approx 2.67$ sub-DLA towards \bgQSO\ and the LLS towards \fgQSO, and (2) the stream around \fgQSO\ that is responsible for the $z_{\rm abs} \approx 2.75$ DLA towards \bgQSO. We have seen the clumpy, filamentary morphology of the streams in the \Lya\ surface brightness maps in Figure\,\ref{fig:moment_maps}, and their connection with the absorption-line systems is evident because the \Lya\ filaments appear to pass in front of their background QSOs. In Figure\,\ref{fig:emission_absorption}, we compare the kinematics of the absorption lines with those of the emission lines. Because both filaments cover a wide range in velocity, for this comparison, we only extract the emission-line spectra from the parts that are closest in projection to the QSO sightlines. The SMG stream offers two comparison positions because there are two background QSOs, while the \fgQSO\ stream offers only one comparison position at \bgQSO. Because the \Lya\ absorption line is highly saturated, we use the \Lyb\ absorption to show the radial velocity distribution of the absorption clouds. For the SMG stream near the position of \bgQSO, the \Lya\ emission is redshifted relative to the sub-DLA (i.e., subsystem A) by $\sim$500\,\kms, positioning it between the two absorption subsystems A and B. On the other hand, near the position of \fgQSO, the emission line is blueshifted by $-$150\,\kms\ relative to the LLS in subsystem A. At both positions, relatively narrow line widths are observed in emission ($\sigma \approx 150$\,\kms) and in absorption \citepalias[$\sigma \approx 30$\,\kms;][Table 4]{Fu21}. For the \fgQSO\ stream, the \Lya\ emission is well aligned with the absorption line and it shows a broad profile with a velocity width of $\sigma \approx 240$\,\kms, which is comparable to the velocity span of the absorption-line clouds traced by the optically thin metal lines \citepalias[see][Figure C2]{Fu21}.

In summary, the comparisons yield a mixed result. The \HI\ \Lya\ emission lines can be either blueshifted or redshifted relative to the \HI\ absorption, but their line widths are comparable to the velocity span covered by the spectroscopically resolved absorption-line clouds. This discrepancy may be attributed to the differences between the two methods to infer gas properties. KCWI measures the average emission over a large area, while absorption lines are narrower pencil-beam measurements.

\subsection{Comparison with \Lya\ Nebulae around Other QSOs} \label{sec:otherQSOs}

\begin{figure}
    \centering
    \hspace*{-0.55cm}  
    \includegraphics[width=0.5\textwidth]{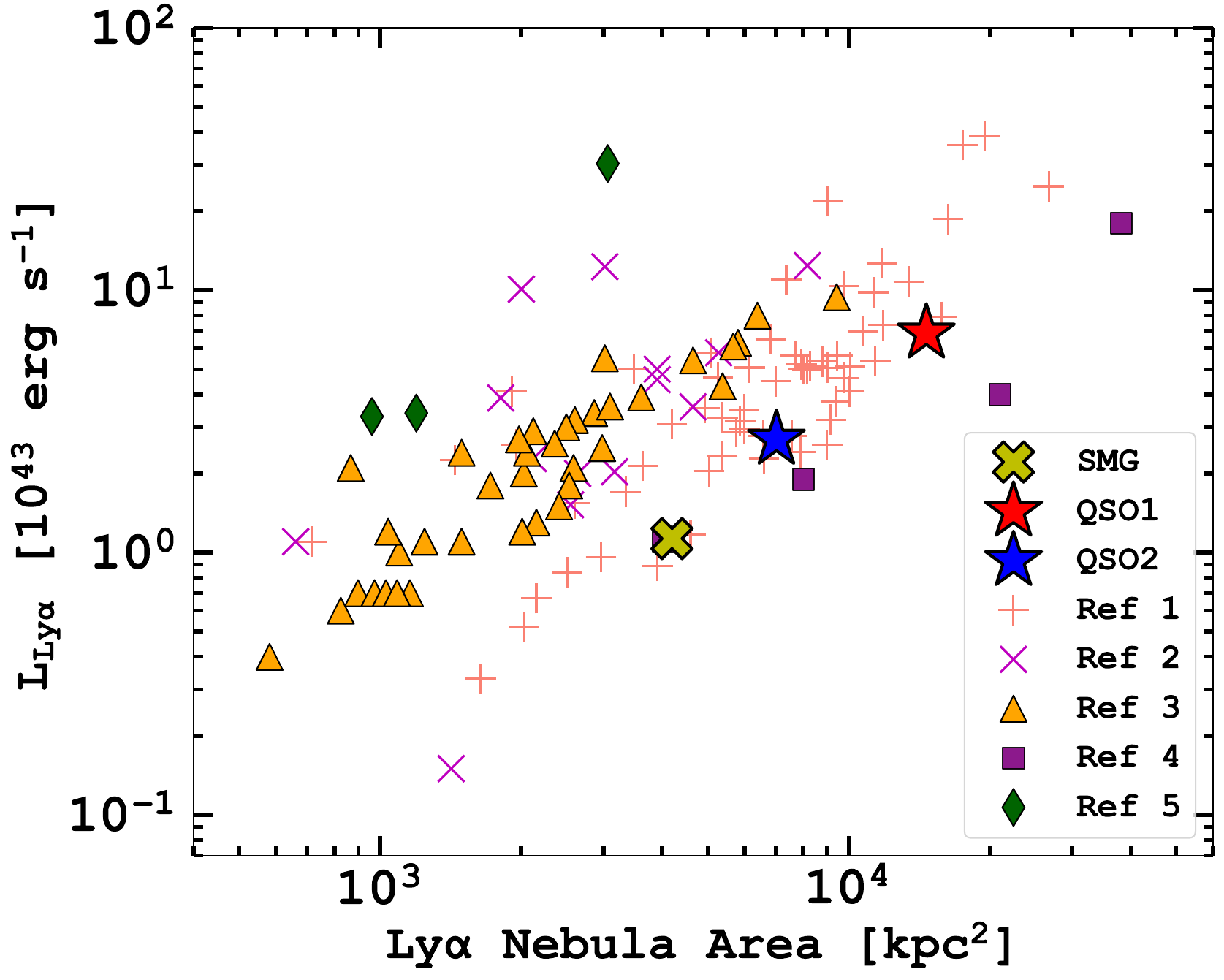}
    \caption{$L_{\rm Ly\alpha}$ as a function of \Lya\ nebula area. The \Lya\ nebula for the SMG, \fgQSO, and \bgQSO\ are plotted as a gold cross, blue star, and red star respectively. The additional datapoints come from previous studies on \Lya\ Nebula, specifically focused on QSO systems. The red crosses correspond to the results from QSO MUSEUM \citep[Ref 1;][]{Arrigoni-Battaia19}, the magenta cross-hairs correspond to \cite{Cai19}'s survey on $z \approx 2$ QSOs with KCWI (Ref 2) , the orange triangles correspond to the QSOs at $z \sim 2.5$ with PCWI \citep[Ref 3;][]{OSullivan20a}. Additionally, we include recent results on the \Lya\ properties of QSO pairs with PCWI \citep{Li23} as purple squares, as well as QSO-SMG composite systems from \cite{Lobos23a} (Ref 5) plotted as green diamonds.}
    \label{fig:neb_comp}
\end{figure}

We will now direct our focus toward comparing the characteristics of each nebula. Figure \ref{fig:neb_comp} facilitates this comparison by juxtaposing our observed nebulae with those reported in the previous studies (QSO nebulae), utilizing the total integrated $L_{\rm Ly\alpha}$ as a function of nebula area. We convert the angular size of each nebula into its physical area. Since we overlay other $z > 2$ QSO results, we begin our comparison with \bgQSO\ and \fgQSO. 

As Figure\,\ref{fig:neb_comp} illustrates, the \Lya\ halos enveloping \bgQSO\ and \fgQSO\ adhere to the established trend between halo $L_{\rm Ly\alpha}$ and nebula area. \citet{Battaia23} employed a similar methodology to establish a Luminosity-Area Relation, which takes the form of $\log(L_{\rm Ly\alpha}^{\rm Neb}) = a_1\log(\rm Area^{\rm Neb}) + a_0$. These relations underscore that the bolometric luminosity ($L_{\rm bol}$) of the QSO housing the \Lya\ nebula does not appear to exert a significant influence on the total $L_{\rm Ly\alpha}$. Our results align with this trend, as both \bgQSO\ and \fgQSO\ exhibit comparable $L_{\rm bol}$ values, despite differing \Lya\ nebula properties. As a final step, we plot the SMG nebula properties and find that it is comparable to other QSO nebulae.

\subsection{Previous Non-Detection of Ly$\alpha$ Nebulae around two SMGs }

Our KCWI observation of the \system\ system presents a rare opportunity to make a direct comparison between the \Lya\ nebula properties for systems hosting SMGs and QSOs. Discovering the nebulae surrounding both \bgQSO\ and \fgQSO\ follows the growing trend of extended \Lya\ emission surrounding $z > 2$ QSOs. However, the same cannot be said for systems hosting SMGs alone. Specifically, we refer to SMGs residing outside of clear protocluster environments where the local UV background may be boosted \citep[e.g.,][]{Umehata19}. The recent study by \cite{Lobos23a} observed SMG-QSO composite systems, as well as SMGs with no QSOs. The authors detect bright \Lya\ nebula within the composite systems, but they do not detect any \Lya\ emission from the CGM of the individual SMGs. This implies that the presence of the QSO could be a dominating mechanism to power the observed nebulae and other mechanisms such as gravitational heating may not be as important.

One critical difference between our KCWI observation of the \system\ system and the MUSE observations performed by \cite{Lobos23a} is the \Lya\ SB detection limit. The authors report a $2\sigma$ limit of $\sim 2\times 10^{-18}$ \surf\ for both SMGs at a redshift of $\approx 4.4$. If we consider cosmological SB dimming and scale this to $z_{\rm SMG} = 2.674$, the limit increases to $\sim 9 \times 10^{-18}$ \surf. Our observations reveal that the nebula reaches a peak \Lya\ of $\sim 5 \times 10^{-18}$\,\surf, nearly $0.5\times$ the limit from the MUSE observations. With this in mind, it is difficult to make a direct comparison.

\section{Summary and Discussion} \label{sec:Summary}

The \system\ system is a rare alignment of three massive $z = 2-3$ galaxies on the sky: one SMG in the foreground of two QSOs. Motivated by previous absorption-line detection of substantial amount of cool gas in their halos, we have carried out and analyzed deep integral-field spectroscopic observations to search for extended \HI\ \Lya\ emission in the halos of the three massive galaxies. Our main findings can be summarized as follows: 

\begin{enumerate}
    \item  We discover a bright \Lya\ nebula surrounding a heavily dust-obscured starburst. The nebula possesses a large filamentary structure stretching $\sim 180$\,kpc from the SMG, intercepting the sightlines of both background QSOs. The full extent of the nebula is $250$\,kpc (31\arcsec), with a total area of $4200$\,kpc$^2$ (67 arcsec$^2$). We find that the peak \Lya\ SB resides near the position of the SMG and reaches a value of $\approx 4.5 \times 10^{-18}$\,\surf\ and measure a total integrated $L_{\rm Ly\alpha}$ of $\approx 1.1 \times 10^{43}$\,erg s$^{-1}$. The velocity map shows that all \Lya\ emission, including the filamentary structure, is consistent with gas that is gravitationally bound to a $10^{13}$\,\msun\ DM halo. 

    \item We detect \Lya\ nebulae surrounding both QSOs at equivalent detection limits and compare their characteristics with the SMG. Both \fgQSO\ and \bgQSO\ have peak \Lya\ SB values that exceed the SMG by $2.7\times$ and $6\times$, respectively. The total $L_{\rm Ly\alpha}$ for each nebula is $\approx 2.7$ and $6.3 \times 10^{43}$\,erg s$^{-1}$. Each nebula has comparable nebula properties with other $z > 2$ QSOs from the literature. The CGM of \fgQSO\ possesses a large filamentary structure stretching $\gtrsim 200$\,kpc from the host galaxy intercepting the sightline of \bgQSO. The DLA towards \bgQSO\ (at $z=2.75$) is attributed to this structure. Furthermore, we find another filament within \bgQSO's nebula that spans nearly $\sim 150$\,kpc to the northwest. Interestingly, the velocity map for \bgQSO\ reveals clear kinematic structure, where we find two regions that have a radial velocity equal to $|\delta v| \approx 300$\,\kms. 
    
    \item We create circularly-averaged SB profiles for each nebula and perform a direct comparison. The SMG has a flatter profile than the QSOs with a power-law index of $\approx -0.7$. Meanwhile, the QSO nebulae exhibit steeper profiles with a power-law index near $\sim -1.8$, which is in agreement with previous $z > 2$ QSO surveys.

    \item Our KCWI observations enable a direct kinematic comparison between the absorption line systems found within each QSO sightline and the associated \Lya\ emission line. The \Lya\ line associated with the sub-DLA (subsystem A) towards \bgQSO\ at $z \approx 2.67$ is redshifted by $\sim +500$\,\kms, but the \Lya\ line is blueshifted by $\sim -150$\,\kms\ with respect to the LLS towards \fgQSO. Both emission lines share a narrow line width of $\sigma \approx 150$\,\kms. For \fgQSO, we find that the \Lya\ emission has a broader line width of $\sigma \approx 240$\,\kms\ and shares equal kinematic properties with its associated DLA system towards \bgQSO. 

\end{enumerate}

The \system\ system is a truly remarkable opportunity to study the extended environments of different high$-z$ galaxies. The two background QSOs enabled a direct study of the SMG's CGM and led to the discovery of a large "shadow" cast by a long filament of gas flowing towards the dusty starburst. Now, with KCWI, we discover the \Lya\ signal emanating from this extended structure. If we consider all of the evidence gathered thus far, it guides us to the conclusion that this is indeed a cool stream fueling the star formation within the SMG, but could there be another scenario that can adequately explain the observations?

One possibility is the role of the obscured AGN/QSO within the dusty starburst ionizing the \Lya\ nebula through {\it unobscured} sightlines. This would not be too dissimilar to the KCWI observation of the SMM J02399-0136 system, which revealed a \Lya-emitting dark cloud powered by an obscured QSO \citep{Li19}. Additionally, luminous \Lya\ nebula have been detected around obscured QSOs as shown by \cite{Brok20} for Type-II AGN. Even though AGN activity is not dominant within the SMG, we still need to consider its role. To do this, we predict the required bolometric luminosity ($L_{\rm Bol}^{\rm required}$) that an obscured QSO would need to power the \Lya\ nebula. This $L_{\rm Bol}^{\rm required}$ value can then be compared with the SMG's total $L_{\rm IR}$ as reported by \citetalias{Fu21} and estimate the escape fraction ($f_{\rm esc}$) of ionizing photons. Additionally, if we find that $L_{\rm Bol}^{\rm required} > L_{\rm IR}$, then we can rule out the possibility of an obscured source.

We begin by utilizing the simple model outlined by \cite{Hennawi13}. The QSO is surrounded by cool ($\sim 10^4$\,K) gas clouds with a uniform volume density ($n_{\rm H}$). From the QSO absorption lines along both sightlines, we know the gas is optically thick \citepalias{Fu21}. As a result, shelf-shielding kicks in and a thin shell surrounds the clouds, converting a fraction of ionizing photons into \Lya\ photons that travel back to the observer. This implies that the \Lya\ SB is proportional to $L_{\nu _{\rm LL}}$, where $L_{\nu _{\rm LL}}$ is the specific luminosity at the Lyman limit for the QSO, and will decrease at greater distances following $R^{-2}$. If we use the observed characteristics of the \Lya\ filament within the CGM ($R \approx 180$\,kpc, $\rm SB_{\rm Ly\alpha} \sim 10^{-18}$\,\surf) and rearrange Equation 12 from \cite{Hennawi13}, we estimate that $\log L_{\nu _{\rm LL}}$ would be $\sim 29.6$\,erg s$^{-1}$ Hz for the {\it unobscured} QSO. 

To get $L_{\rm Bol}^{\rm required}$, we must perform a bolometric correction. We convert $\log L_{\nu _{\rm LL}}$ to the rest-frame $1350$\,\AA\ luminosity ($L_{1350}$) by assuming the QSO's SED follows the power-law: $L_{\nu} = L_{\nu_{\rm LL}}(\nu / \nu_{\rm LL})^{-1.7}$ \citep{Lusso15}. Next, we perform the bolometric correction on to $L_{1350}$ following Equation 10 in \cite{Fu17}, which utilizes the QSO SEDs from \cite{Hopkins07}. We find that $L_{\rm Bol}^{\rm required} \approx 1.7\times10^{12}$\,\lsun. Finally, we estimate $f_{\rm esc}$ by taking the ratio $L_{\rm Bol}^{\rm required} / L_{\rm IR}$ and compute a value of $\sim 15\%$. This is larger than expected for the SMG if we consider its dust content. However, it is still within a plausible limit of the true $f_{\rm esc}$ value because of the large uncertainty in our calculation due to the assumptions used. Furthermore, this estimate did not consider any possible local enhancements to the ionizing background in which the SMG resides. As a result, it is unclear from this estimate whether an obscured source can explain our observations, but it can likely be excluded due to the lack of AGN activity within the SMG.

We also consider the role of obscured star formation within the nebula. First, the SMG is heavily obscured. This can be appreciated by the undetected UV continuum in the deep KiDS image (Figure \ref{fig:finderchart}), as well as the dust properties measured by ALMA \citepalias{Fu21}. Assuming all \Lya\ photons are produced by the internal SFR under Case-B recombination\footnote{We utilize Equation 2 from \cite{Robert-C.-Kennicutt98} and convert the SFR$_{\rm IR}$ to SFR$_{\rm UV}$ \citep{Kennicutt12} assuming a constant star-formation history.}, we would expect a total $L_{\rm Ly\alpha}$ of $\sim 1.2 \times 10^{44}$\,erg s$^{-1}$. This would require $f_{\rm esc} \sim 10\%$, which is similar to the previous estimate. Again, this is likely higher than the true value for the SMG due to the lack of a detected continuum.

Furthermore, several \Lya\ nebulae are found to host dusty star forming galaxies \citep{Geach07,Geach16,Oteo18,Umehata21}. However, these bright nebulae are host to several galaxies within a relatively small physical size. \cite{Umehata21} identified 9 galaxies within LAB1 and measured a total SFR $\approx 200$\,\msunyr. If we compare the \Lya\ properties of LAB1 and other filamentary emission associated with SMGs from the SSA22 protocluser \citep{Umehata19}, we find that the \Lya\ SB is noticeably lower around the SMGs (with no associated AGN). If obscured star formation plays a dominant role, we would expect these isolated SMGs to display either equal or higher \Lya\ SB since they possess higher rates of star formation \citep{Umehata17}. \cite{Umehata21} suggests an additional mechanism may be needed to explain the \Lya\ nebula, such as cooling flows or galaxy mergers. The observations by \cite{Lobos23a} also support this conclusion. If star formation is the dominant mechanism, then the authors should have detected \Lya\ nebulae around their SMGs at their detection level. However, this argument assumes a homogeneous environment surrounding SMGs such as the cold gas density and clumpiness. As a result, these comparisons alone cannot eliminate the role of star-formation.

\citetalias{Fu21} utilized UV background radiation models from \cite{Haardt12} (interpolated to $z=2.67$) to act as the ionizing source in their photoionization models. However, if this is the sole ionizing source, we would expect a mean \Lya\ SB of the order $\sim 10^{-20}$\,\surf\ \citep{Gould96}, which is an order of magnitude lower than our detection limit. However, the UV background may be boosted by nearby star forming galaxies or AGN. Indeed, overdense regions, such as the environment of the SMG, can possibly supply the necessary photon budget to power the observed \Lya\ nebula. It has also been shown that overdensities of LAEs are associated with bright \Lya\ nebulae \citep{Matsuda11, Zhang23a}, and the identification of LAE1 near the SMG may suggest a similarity. However, it is unclear whether the \system\ system represents a newly discovered protocluster region, such as the observations presented by \cite{Umehata19}. In the ALMA \cothree\ \citepalias{Fu21} and 870\,\um\ \citep{Fu17} maps, it was confirmed that the bright submillimeter source was not a blend of multiple interacting galaxies, which was the case for LAB1 early on \citep{Matsuda07,Geach14}. With KCWI, we do not identify any additionally galaxies near the SMG outside of LAE1. 

Despite the uncertainty in classifying the environment of the SMG as a protocluster, we can be confident that it does indeed reside within an overdense environment. The identification of other CO emitters within 30\arcsec\ of the SMG would require a survey size that covers $\sim 10\times$ the area for detections of companions \citepalias{Fu21}. This estimated overdensity is comparable to one of the largest overdensities of CO emitters \citep{Pensabene24}. This supports the possibility of the \Lya\ stream being photoionized by a UV background higher than the models from \cite{Haardt12}. Nevertheless, future observations at larger scales will be needed to better constrain the UV background. 

We now consider the role of gravitational energy. The gas will dissipate gravitational potential energy during infall and form shocks as they flow closer to the starburst. Cosmological hydrodynamical simulations predict \Lya\ nebulae with $L_{\rm Ly\alpha} \approx 10^{43}-10^{44}$\,erg s$^{-1}$ for $\sim 10^{12-13}$\,\msun\ DM halos \citep{Faucher-Giguere10,Rosdahl12}, which is consistent with our measurements for the SMG. However, these predictions are highly uncertain, as local boosts in UV photons from embedded star formation can increase the expected $L_{\rm Ly\alpha}$. Specifically, LAE1 likely contributes \Lya\ photons that can ionize regions of the nebula, as it is an unobscured source. Interestingly, the $\pm 300$\,\kms\ velocity gradient detected $\sim 5$\,arcsecs to the East of the SMG may also indicate gas infall from a second filament that is cutoff from our KCWI FOV. Similar velocity structure was detected by \cite{Daddi21} for the RO-1001 group at $z=2.91$. Furthermore, the SMG resides at the region of highest surface brightness. This indicates that the nebulae is at the center of mass of the DM halo, which is an expectation if gravity plays a role. Furthermore, the \Lya\ filament itself, if we interpret as a cooling flow, is an indicator for the influence of gravity. The filament may also be ionized by the surrounding hot CGM as it flows towards the starburst. Overall, the evidence provided by the absorption study and the KCWI observations suggests cool gas accretion can serve as a plausible scenario. 

The results presented here demonstrates the advantage of tying together QSO absorption line spectroscopy probing the CGM of SMGs with deep optical IFU observations. The presence of the optically thick absorbers towards both QSOs suggested that the SMG holds exciting dynamics within its CGM. More importantly, it provided the metallicity of the gas and excluded the possibility of metal enriched outflows. Searching for other conjunctions between QSOs and foreground SMGs can aide our effort for further observational evidence for cool streams flowing through hot CGMs, as well as help us prioritize future targets. Additionally, searching for spatially extended \Lya\ emission within the trough of the DLA at $z=2.75$ towards \bgQSO\ revealed an unexpectedly large extended \Lya\ filament within the CGM of \fgQSO. Could filamentary structure be a characteristic of CGMs that possess DLAs? The primary challenge we face is the rarity of the \system\ system. We will need to search for more QSO constellations with foreground SMGs. Utilizing the latest data releases from large sky surveys, such as the Dark Energy Spectroscopic Instrument (DESI), can aide our search and enable future discoveries. Similar to the ongoing QSO \Lya\ halo studies, performing large IFU surveys on SMGs, similar to the one studied here, can help us better understand the CGM of these extreme starbursts.

\acknowledgments
We thank the anonymous referee for their detailed comments that helped improve the paper.
This work is supported by the National Science Foundation (NSF) grants AST-1614326 and AST-2103251.
The authors wish to recognize and acknowledge the very significant cultural role and reverence that the summit of Maunakea has always had within the indigenous Hawaiian community.  We are most fortunate to have the opportunity to conduct observations from this mountain.
All analysis software and data used in this study can be found on the corresponding author's Github page\footnote{\url{https://github.com/kevhall23}}

{\it Facilities}: Keck/KCWI, KiDS

\bibliographystyle{apj}
\bibliography{kcwi}

\begin{thebibliography}{93}
\expandafter\ifx\csname natexlab\endcsname\relax\def\natexlab#1{#1}\fi

\bibitem[{Alexander {et~al.}(2005)Alexander, Bauer, Chapman, Smail, Blain,
  Brandt, \& Ivison}]{Alexander05}
Alexander, D.~M., Bauer, F.~E., Chapman, S.~C., {et~al.} 2005, \apj, 632, 736

\bibitem[{{Arrigoni Battaia} {et~al.}(2019){Arrigoni Battaia}, {Hennawi},
  {Prochaska}, {O{\~n}orbe}, {Farina}, {Cantalupo}, \&
  {Lusso}}]{Arrigoni-Battaia19}
{Arrigoni Battaia}, F., {Hennawi}, J.~F., {Prochaska}, J.~X., {et~al.} 2019,
  \mnras, 482, 3162

\bibitem[{Bacon {et~al.}(2010)Bacon, Accardo, Adjali, Anwand, Bauer, Biswas,
  Blaizot, Boudon, Brau-Nogue, Brinchmann, Caillier, Capoani, Carollo, Contini,
  Couderc, Daguis{\'e}, Deiries, Delabre, Dreizler, Dubois, Dupieux, Dupuy,
  Emsellem, Fechner, Fleischmann, Fran{\c c}ois, Gallou, Gharsa, Glindemann,
  Gojak, Guiderdoni, Hansali, Hahn, Jarno, Kelz, Koehler, Kosmalski, Laurent,
  Le~Floch, Lilly, Lizon, Loupias, Manescau, Monstein, Nicklas, Olaya, Pares,
  Pasquini, P{\'e}contal-Rousset, Pell{\'o}, Petit, Popow, Reiss, Remillieux,
  Renault, Roth, Rupprecht, Serre, Schaye, Soucail, Steinmetz, Streicher,
  Stuik, Valentin, Vernet, Weilbacher, Wisotzki, \& Yerle}]{Bacon10}
Bacon, R., Accardo, M., Adjali, L., {et~al.} 2010, SPIE

\bibitem[{Barger {et~al.}(1998)Barger, Cowie, Sanders, Fulton, Taniguchi, Sato,
  Kawara, \& Okuda}]{Barger98}
Barger, A.~J., Cowie, L.~L., Sanders, D.~B., {et~al.} 1998, Nature, 394, 248

\bibitem[{Battaia {et~al.}(2023)Battaia, Obreja, Costa, Farina, \&
  Cai}]{Battaia23}
Battaia, F.~A., Obreja, A., Costa, T., Farina, E.~P., \& Cai, Z. 2023, \apjl,
  952, L24

\bibitem[{Berk {et~al.}(2001)Berk, Richards, Bauer, Strauss, Schneider,
  Heckman, York, Hall, Fan, Knapp, Anderson, Annis, Bahcall, Bernardi, Briggs,
  Brinkmann, Brunner, Burles, Carey, Castander, Connolly, Crocker, Csabai, Doi,
  Finkbeiner, Friedman, Frieman, Fukugita, Gunn, Hennessy, Ivezi{\'{c}}, Kent,
  Kunszt, Lamb, Leger, Long, Loveday, Lupton, Meiksin, Merelli, Munn, Newberg,
  Newcomb, Nichol, Owen, Pier, Pope, Rockosi, Schlegel, Siegmund, Smee, Snir,
  Stoughton, Stubbs, SubbaRao, Szalay, Szokoly, Tremonti, Uomoto, Waddell,
  Yanny, \& Zheng}]{Berk01}
Berk, D. E.~V., Richards, G.~T., Bauer, A., {et~al.} 2001, \aj, 122, 549

\bibitem[{Berry {et~al.}(2012)Berry, Gawiser, Guaita, Padilla, Treister, Blanc,
  Ciardullo, Francke, \& Gronwall}]{Berry12}
Berry, M., Gawiser, E., Guaita, L., {et~al.} 2012, \apj, 749, 4

\bibitem[{{Birnboim} \& {Dekel}(2003)}]{Birnboim03}
{Birnboim}, Y., \& {Dekel}, A. 2003, \mnras, 345, 349

\bibitem[{Blain(2002)}]{Blain02}
Blain, A. 2002, Physics Reports, 369, 111

\bibitem[{{Borisova} {et~al.}(2016){Borisova}, {Cantalupo}, {Lilly}, {Marino},
  {Gallego}, {Bacon}, {Blaizot}, {Bouch{\'e}}, {Brinchmann}, {Carollo},
  {Caruana}, {Finley}, {Herenz}, {Richard}, {Schaye}, {Straka}, {Turner},
  {Urrutia}, {Verhamme}, \& {Wisotzki}}]{Borisova16}
{Borisova}, E., {Cantalupo}, S., {Lilly}, S.~J., {et~al.} 2016, \apj, 831, 39

\bibitem[{Byrohl \& Nelson(2022)}]{Byrohl22}
Byrohl, C., \& Nelson, D. 2022, arXiv

\bibitem[{{Cai} {et~al.}(2017){Cai}, {Fan}, {Yang}, {Bian}, {Prochaska},
  {Zabludoff}, {McGreer}, {Zheng}, {Green}, {Cantalupo}, {Frye}, {Hamden},
  {Jiang}, {Kashikawa}, \& {Wang}}]{Cai17}
{Cai}, Z., {Fan}, X., {Yang}, Y., {et~al.} 2017, \apj, 837, 71

\bibitem[{{Cai} {et~al.}(2018){Cai}, {Hamden}, {Matuszewski}, {Prochaska},
  {Li}, {Cantalupo}, {Arrigoni Battaia}, {Martin}, {Neill}, {O'Sullivan},
  {Wang}, {Moore}, \& {Morrissey}}]{Cai18}
{Cai}, Z., {Hamden}, E., {Matuszewski}, M., {et~al.} 2018, \apjl, 861, L3

\bibitem[{{Cai} {et~al.}(2019){Cai}, {Cantalupo}, {Prochaska}, {Arrigoni
  Battaia}, {Burchett}, {Li}, {Chisholm}, {Bundy}, \& {Hennawi}}]{Cai19}
{Cai}, Z., {Cantalupo}, S., {Prochaska}, J.~X., {et~al.} 2019, \apjs, 245, 23

\bibitem[{{Cantalupo} {et~al.}(2014){Cantalupo}, {Arrigoni-Battaia},
  {Prochaska}, {Hennawi}, \& {Madau}}]{Cantalupo14}
{Cantalupo}, S., {Arrigoni-Battaia}, F., {Prochaska}, J.~X., {Hennawi}, J.~F.,
  \& {Madau}, P. 2014, \nat, 506, 63

\bibitem[{Cantalupo {et~al.}(2012)Cantalupo, Lilly, \& Haehnelt}]{Cantalupo12}
Cantalupo, S., Lilly, S.~J., \& Haehnelt, M.~G. 2012, \mnras, 425, 1992

\bibitem[{{Cantalupo} {et~al.}(2005){Cantalupo}, {Porciani}, {Lilly}, \&
  {Miniati}}]{Cantalupo05}
{Cantalupo}, S., {Porciani}, C., {Lilly}, S.~J., \& {Miniati}, F. 2005, \apj,
  628, 61

\bibitem[{Cen \& Zheng(2013)}]{Cen13}
Cen, R., \& Zheng, Z. 2013, \apj, 775, 112

\bibitem[{Chapman {et~al.}(2005)Chapman, Blain, Smail, \& Ivison}]{Chapman05}
Chapman, S.~C., Blain, A.~W., Smail, I., \& Ivison, R.~J. 2005, \apj, 622, 772

\bibitem[{Chapman {et~al.}(2001)Chapman, Lewis, Scott, Richards, Borys,
  Steidel, Adelberger, \& Shapley}]{Chapman01}
Chapman, S.~C., Lewis, G.~F., Scott, D., {et~al.} 2001, \apj, 548, L17

\bibitem[{Da{\^{A}}ngela {et~al.}(2007)Da{\^{A}}ngela, Shanks, Croom,
  Weilbacher, Brunner, Couch, Miller, Myers, Nichol, Pimbblet, Propris,
  Richards, Ross, Schneider, \& Wake}]{DaAngela07}
Da{\^{A}}ngela, J., Shanks, T., Croom, S.~M., {et~al.} 2007, \mnras, 383, 565

\bibitem[{{Daddi} {et~al.}(2021){Daddi}, {Valentino}, {Rich}, {Neill},
  {Gronke}, {O'Sullivan}, {Elbaz}, {Bournaud}, {Finoguenov}, {Marchal},
  {Delvecchio}, {Jin}, {Liu}, {Strazzullo}, {Calabro}, {Coogan}, {D'Eugenio},
  {Gobat}, {Kalita}, {Laursen}, {Martin}, {Puglisi}, {Schinnerer}, \&
  {Wang}}]{Daddi21}
{Daddi}, E., {Valentino}, F., {Rich}, R.~M., {et~al.} 2021, \aap, 649, A78

\bibitem[{{de Jong} {et~al.}(2013){de Jong}, {Verdoes Kleijn}, {Kuijken}, \&
  {Valentijn}}]{de-Jong13}
{de Jong}, J. T.~A., {Verdoes Kleijn}, G.~A., {Kuijken}, K.~H., \& {Valentijn},
  E.~A. 2013, Experimental Astronomy, 35, 25

\bibitem[{{Dekel} \& {Birnboim}(2006)}]{Dekel06}
{Dekel}, A., \& {Birnboim}, Y. 2006, \mnras, 368, 2

\bibitem[{{Dekel} {et~al.}(2009){Dekel}, {Birnboim}, {Engel}, {Freundlich},
  {Goerdt}, {Mumcuoglu}, {Neistein}, {Pichon}, {Teyssier}, \&
  {Zinger}}]{Dekel09}
{Dekel}, A., {Birnboim}, Y., {Engel}, G., {et~al.} 2009, \nat, 457, 451

\bibitem[{den Brok {et~al.}(2020)den Brok, Cantalupo, Mackenzie, Marino,
  Pezzulli, Matthee, Johnson, Krumpe, Urrutia, \& Kollatschny}]{Brok20}
den Brok, J.~S., Cantalupo, S., Mackenzie, R., {et~al.} 2020, MNRAS, 495, 1874

\bibitem[{Dey {et~al.}(2005)Dey, Bian, Soifer, Brand, Brown, Chaffee,
  Le~Floc'h, Hill, Houck, Jannuzi, Rieke, Weedman, Brodwin, \&
  Eisenhardt}]{Dey05}
Dey, A., Bian, C., Soifer, B.~T., {et~al.} 2005, \apj, 629, 654

\bibitem[{{Dijkstra} \& {Loeb}(2009)}]{Dijkstra09}
{Dijkstra}, M., \& {Loeb}, A. 2009, \mnras, 400, 1109

\bibitem[{{Drake} {et~al.}(2020){Drake}, {Walter}, {Novak}, {Farina},
  {Neeleman}, {Riechers}, {Carilli}, {Decarli}, {Mazzucchelli}, \&
  {Onoue}}]{Drake20}
{Drake}, A.~B., {Walter}, F., {Novak}, M., {et~al.} 2020, \apj, 902, 37

\bibitem[{Farina {et~al.}(2019)Farina, Arrigoni-Battaia, Costa, Walter,
  Hennawi, Drake, Decarli, Gutcke, Mazzucchelli, Neeleman, Georgiev, Eilers,
  Davies, Ba{\~{n}}ados, Fan, Onoue, Schindler, Venemans, Wang, Yang, Rabien,
  \& Busoni}]{Farina19}
Farina, E.~P., Arrigoni-Battaia, F., Costa, T., {et~al.} 2019, \apj, 887, 196

\bibitem[{Faucher-Gigu{\`{e}}re {et~al.}(2011)Faucher-Gigu{\`{e}}re,
  Kere{\v{s}}, \& Ma}]{Faucher-Giguere11}
Faucher-Gigu{\`{e}}re, C.-A., Kere{\v{s}}, D., \& Ma, C.-P. 2011, \mnras, 417,
  2982

\bibitem[{{Faucher-Gigu{\`e}re} {et~al.}(2010){Faucher-Gigu{\`e}re},
  {Kere{\v{s}}}, {Dijkstra}, {Hernquist}, \& {Zaldarriaga}}]{Faucher-Giguere10}
{Faucher-Gigu{\`e}re}, C.-A., {Kere{\v{s}}}, D., {Dijkstra}, M., {Hernquist},
  L., \& {Zaldarriaga}, M. 2010, \apj, 725, 633

\bibitem[{{Fossati} {et~al.}(2021){Fossati}, {Fumagalli}, {Lofthouse}, {Dutta},
  {Cantalupo}, {Arrigoni Battaia}, {Fynbo}, {Lusso}, {Murphy}, {Prochaska},
  {Theuns}, \& {Cooke}}]{Fossati21}
{Fossati}, M., {Fumagalli}, M., {Lofthouse}, E.~K., {et~al.} 2021, \mnras, 503,
  3044

\bibitem[{Fruchter \& Hook(2002)}]{Fruchter02}
Fruchter, A.~S., \& Hook, R.~N. 2002, \pasp, 114, 144

\bibitem[{{Fu} {et~al.}(2017){Fu}, {Isbell}, {Casey}, {Cooray}, {Prochaska},
  {Scoville}, \& {Stockton}}]{Fu17}
{Fu}, H., {Isbell}, J., {Casey}, C.~M., {et~al.} 2017, \apj, 844, 123

\bibitem[{{Fu} {et~al.}(2021){Fu}, {Xue}, {Prochaska}, {Stockton}, {Ponnada},
  {Lau}, {Cooray}, \& {Narayanan}}]{Fu21}
{Fu}, H., {Xue}, R., {Prochaska}, J.~X., {et~al.} 2021, \apj, 908, 188

\bibitem[{{Fu} {et~al.}(2016){Fu}, {Hennawi}, {Prochaska}, {Mutel}, {Casey},
  {Cooray}, {Kere{\v{s}}}, {Zhang}, {Clements}, {Isbell}, {Lang}, {McGinnis},
  {Micha{\l}owski}, {Mooley}, {Perley}, {Stockton}, \& {Thompson}}]{Fu16}
{Fu}, H., {Hennawi}, J.~F., {Prochaska}, J.~X., {et~al.} 2016, \apj, 832, 52

\bibitem[{Geach {et~al.}(2007)Geach, Smail, Chapman, Alexander, Blain, Stott,
  \& Ivison}]{Geach07}
Geach, J.~E., Smail, I., Chapman, S.~C., {et~al.} 2007, \apj, 655, L9

\bibitem[{Geach {et~al.}(2014)Geach, Bower, Alexander, Blain, Bremer, Chapin,
  Chapman, Clements, Coppin, Dunlop, Farrah, Jenness, Koprowski,
  Micha{\l}owski, Robson, Scott, Smith, Spaans, Swinbank, \& van~der
  Werf}]{Geach14}
Geach, J.~E., Bower, R.~G., Alexander, D.~M., {et~al.} 2014, \apj, 793, 22

\bibitem[{Geach {et~al.}(2016)Geach, Narayanan, Matsuda, Hayes, Mas-Ribas,
  Dijkstra, Steidel, Chapman, Feldmann, Avison, Agertz, Ao, Birkinshaw, Bremer,
  Clements, Dannerbauer, Farrah, Harrison, Kubo, Micha{\l}owski, Scott, Smith,
  Spaans, Simpson, Swinbank, Taniguchi, Werf, Verma, \& Yamada}]{Geach16}
Geach, J.~E., Narayanan, D., Matsuda, Y., {et~al.} 2016, \apj, 832, 37

\bibitem[{Goerdt {et~al.}(2010)Goerdt, Dekel, Sternberg, Ceverino, Teyssier, \&
  Primack}]{Goerdt10}
Goerdt, T., Dekel, A., Sternberg, A., {et~al.} 2010, \mnras, 407, 613

\bibitem[{Gould \& Weinberg(1996)}]{Gould96}
Gould, A., \& Weinberg, D.~H. 1996, \apj, 468, 462

\bibitem[{Gronke {et~al.}(2017)Gronke, Dijkstra, McCourt, \& Oh}]{Gronke17}
Gronke, M., Dijkstra, M., McCourt, M., \& Oh, S.~P. 2017, \aap, 607, A71

\bibitem[{{Haardt} \& {Madau}(2012)}]{Haardt12}
{Haardt}, F., \& {Madau}, P. 2012, \apj, 746, 125

\bibitem[{{Hennawi} \& {Prochaska}(2013)}]{Hennawi13}
{Hennawi}, J.~F., \& {Prochaska}, J.~X. 2013, \apj, 766, 58

\bibitem[{Hennawi {et~al.}(2015)Hennawi, Prochaska, Cantalupo, \&
  Arrigoni-Battaia}]{Hennawi15}
Hennawi, J.~F., Prochaska, J.~X., Cantalupo, S., \& Arrigoni-Battaia, F. 2015,
  Science, 348, 779

\bibitem[{Hennawi {et~al.}(2006)Hennawi, Prochaska, Burles, Strauss, Richards,
  Schlegel, Fan, Schneider, Zakamska, Oguri, Gunn, Lupton, \&
  Brinkmann}]{Hennawi06}
Hennawi, J.~F., Prochaska, J.~X., Burles, S., {et~al.} 2006, \apj, 651, 61

\bibitem[{Hickox {et~al.}(2012)Hickox, Wardlow, Smail, Myers, Alexander,
  Swinbank, Danielson, Stott, Chapman, Coppin, Dunlop, Gawiser, Lutz, van~der
  Werf, \& Wei{\ss}}]{Hickox12}
Hickox, R.~C., Wardlow, J.~L., Smail, I., {et~al.} 2012, \mnras, no

\bibitem[{Hopkins {et~al.}(2007)Hopkins, Richards, \& Hernquist}]{Hopkins07}
Hopkins, P.~F., Richards, G.~T., \& Hernquist, L. 2007, \apj, 654, 731

\bibitem[{Jones {et~al.}(2012)Jones, Stark, \& Ellis}]{Jones12}
Jones, T., Stark, D.~P., \& Ellis, R.~S. 2012, \apj, 751, 51

\bibitem[{Kennicutt \& Evans(2012)}]{Kennicutt12}
Kennicutt, R.~C., \& Evans, N.~J. 2012, Annual Review of Astronomy and
  Astrophysics, 50, 531

\bibitem[{{Kere{\v{s}}} {et~al.}(2009){Kere{\v{s}}}, {Katz}, {Dav{\'e}},
  {Fardal}, \& {Weinberg}}]{Keres09}
{Kere{\v{s}}}, D., {Katz}, N., {Dav{\'e}}, R., {Fardal}, M., \& {Weinberg},
  D.~H. 2009, \mnras, 396, 2332

\bibitem[{{Kere{\v{s}}} {et~al.}(2005){Kere{\v{s}}}, {Katz}, {Weinberg}, \&
  {Dav{\'e}}}]{Keres05}
{Kere{\v{s}}}, D., {Katz}, N., {Weinberg}, D.~H., \& {Dav{\'e}}, R. 2005,
  \mnras, 363, 2

\bibitem[{Kesseli {et~al.}(2017)Kesseli, West, Veyette, Harrison, Feldman, \&
  Bochanski}]{Kesseli17}
Kesseli, A.~Y., West, A.~A., Veyette, M., {et~al.} 2017, \apjs, 230, 16

\bibitem[{{Kollmeier} {et~al.}(2010){Kollmeier}, {Zheng}, {Dav{\'e}}, {Gould},
  {Katz}, {Miralda-Escud{\'e}}, \& {Weinberg}}]{Kollmeier10}
{Kollmeier}, J.~A., {Zheng}, Z., {Dav{\'e}}, R., {et~al.} 2010, \apj, 708, 1048

\bibitem[{Lau {et~al.}(2022)Lau, Hamann, Gillette, Perrotta, Rupke, Wylezalek,
  \& Zakamska}]{Lau22}
Lau, M.~W., Hamann, F., Gillette, J., {et~al.} 2022, \mnras, 515, 1624

\bibitem[{Li {et~al.}(2023)Li, Vargas, O'Sullivan, Hamden, Cai, Matuszewski,
  Martin, Keppler, Chung, Melso, \& Zhang}]{Li23}
Li, J.~S., Vargas, C.~J., O'Sullivan, D., {et~al.} 2023, \apj, 952, 137

\bibitem[{{Li} {et~al.}(2019){Li}, {Cai}, {Prochaska}, {Arrigoni Battaia},
  {Ivison}, {Falgarone}, {Cantalupo}, {Matuszewski}, {Neill}, {Wang}, {Martin},
  \& {Moore}}]{Li19}
{Li}, Q., {Cai}, Z., {Prochaska}, J.~X., {et~al.} 2019, \apj, 875, 130

\bibitem[{Lobos {et~al.}(2023)Lobos, Battaia, Chang, Gronke, Kauffmann, Chen,
  Fu, Obreja, \& Farina}]{Lobos23a}
Lobos, V.~G., Battaia, F.~A., Chang, S.-J., {et~al.} 2023, arXiv

\bibitem[{Lusso {et~al.}(2015)Lusso, Worseck, Hennawi, Prochaska, Vignali,
  Stern, \& O'Meara}]{Lusso15}
Lusso, E., Worseck, G., Hennawi, J.~F., {et~al.} 2015, MNRAS, 449, 4204

\bibitem[{Mandelker {et~al.}(2019)Mandelker, Nagai, Aung, Dekel, Padnos, \&
  Birnboim}]{Mandelker19}
Mandelker, N., Nagai, D., Aung, H., {et~al.} 2019, \mnras, 484, 1100

\bibitem[{{Martin} {et~al.}(2019){Martin}, {O'Sullivan}, {Matuszewski},
  {Hamden}, {Dekel}, {Lapiner}, {Morrissey}, {Neill}, {Cantalupo}, {Prochaska},
  {Steidel}, {Trainor}, {Moore}, {Ceverino}, {Primack}, \& {Rizzi}}]{Martin19}
{Martin}, D.~C., {O'Sullivan}, D., {Matuszewski}, M., {et~al.} 2019, Nature
  Astronomy, 3, 822

\bibitem[{Matsuda {et~al.}(2007)Matsuda, Iono, Ohta, Yamada, Kawabe, Hayashino,
  Peck, \& Petitpas}]{Matsuda07}
Matsuda, Y., Iono, D., Ohta, K., {et~al.} 2007, \apj, 667, 667

\bibitem[{Matsuda {et~al.}(2004)Matsuda, Yamada, Hayashino, Tamura, Yamauchi,
  Ajiki, Fujita, Murayama, Nagao, Ohta, Okamura, Ouchi, Shimasaku, Shioya, \&
  Taniguchi}]{Matsuda04}
Matsuda, Y., Yamada, T., Hayashino, T., {et~al.} 2004, \aj, 128, 569

\bibitem[{Matsuda {et~al.}(2011)Matsuda, Yamada, Hayashino, Yamauchi, Nakamura,
  Morimoto, Ouchi, Ono, Kousai, Nakamura, Horie, Fujii, Umemura, \&
  Mori}]{Matsuda11}
---. 2011, MNRAS, 410, L13

\bibitem[{{Morrissey} {et~al.}(2018){Morrissey}, {Matuszewski}, {Martin},
  {Neill}, {Epps}, {Fucik}, {Weber}, {Darvish}, {Adkins}, {Allen}, {Bartos},
  {Belicki}, {Cabak}, {Callahan}, {Cowley}, {Crabill}, {Deich}, {Delecroix},
  {Doppman}, {Hilyard}, {James}, {Kaye}, {Kokorowski}, {Kwok}, {Lanclos},
  {Milner}, {Moore}, {O'Sullivan}, {Parihar}, {Park}, {Phillips}, {Rizzi},
  {Rockosi}, {Rodriguez}, {Salaun}, {Seaman}, {Sheikh}, {Weiss}, \&
  {Zarzaca}}]{Morrissey18}
{Morrissey}, P., {Matuszewski}, M., {Martin}, D.~C., {et~al.} 2018, \apj, 864,
  93

\bibitem[{Neill {et~al.}(2018)Neill, Development, \& mattphys}]{neill18}
Neill, D., Development, K., \& mattphys. 2018,
  {Keck-DataReductionPipelines/KcwiDRP: KCWI Data Reduction Pipeline: First
  minor release}

\bibitem[{Nelson {et~al.}(2015)Nelson, Genel, Vogelsberger, Springel, Sijacki,
  Torrey, \& Hernquist}]{Nelson15}
Nelson, D., Genel, S., Vogelsberger, M., {et~al.} 2015, \mnras, 448, 59

\bibitem[{Nilsson {et~al.}(2006)Nilsson, Fynbo, M{\o}ller, Sommer-Larsen, \&
  Ledoux}]{Nilsson06}
Nilsson, K.~K., Fynbo, J. P.~U., M{\o}ller, P., Sommer-Larsen, J., \& Ledoux,
  C. 2006, \aap, 452, L23

\bibitem[{{O'Sullivan} \& {Chen}(2020)}]{OSullivan20}
{O'Sullivan}, D., \& {Chen}, Y. 2020, arXiv e-prints, arXiv:2011.05444

\bibitem[{{O'Sullivan} {et~al.}(2020){O'Sullivan}, {Martin}, {Matuszewski},
  {Hoadley}, {Hamden}, {Neill}, {Lin}, \& {Parihar}}]{OSullivan20a}
{O'Sullivan}, D.~B., {Martin}, C., {Matuszewski}, M., {et~al.} 2020, \apj, 894,
  3

\bibitem[{Oteo {et~al.}(2018)Oteo, Ivison, Dunne, Manilla-Robles, Maddox,
  Lewis, Zotti, Bremer, Clements, Cooray, Dannerbauer, Eales, Greenslade,
  Omont, Perez-Fourn{\'o}n, Riechers, Scott, van~der Werf, Weiss, \&
  Zhang}]{Oteo18}
Oteo, I., Ivison, R.~J., Dunne, L., {et~al.} 2018, \apj, 856, 72

\bibitem[{Pensabene {et~al.}(2024)Pensabene, Cantalupo, Cicone, Decarli,
  Galbiati, Ginolfi, de~Beer, Fossati, Fumagalli, Lazeyras, Pezzulli,
  Travascio, Wang, Matthee, \& Maseda}]{Pensabene24}
Pensabene, A., Cantalupo, S., Cicone, C., {et~al.} 2024, ALMA survey of a
  massive node of the Cosmic Web at z~3. I. Discovery of a large overdensity of
  CO emitters

\bibitem[{Prochaska {et~al.}(2013)Prochaska, Hennawi, Lee, Cantalupo, Bovy,
  Djorgovski, Ellison, Lau, Martin, Myers, Rubin, \& Simcoe}]{Prochaska13}
Prochaska, J.~X., Hennawi, J.~F., Lee, K.-G., {et~al.} 2013, \apj, 776, 136

\bibitem[{Rees \& Ostriker(1977)}]{Rees77}
Rees, M.~J., \& Ostriker, J.~P. 1977, \mnras, 179, 541

\bibitem[{Robert C.~Kennicutt(1998)}]{Robert-C.-Kennicutt98}
Robert C.~Kennicutt, J. 1998, \apj, 498, 541

\bibitem[{{Rosdahl} \& {Blaizot}(2012)}]{Rosdahl12}
{Rosdahl}, J., \& {Blaizot}, J. 2012, \mnras, 423, 344

\bibitem[{Shapley {et~al.}(2003)Shapley, Steidel, Pettini, \&
  Adelberger}]{Shapley03}
Shapley, A.~E., Steidel, C.~C., Pettini, M., \& Adelberger, K.~L. 2003, \apj,
  588, 65

\bibitem[{Silk(1977)}]{Silk77}
Silk, J. 1977, \apj, 211, 638

\bibitem[{{Smail} {et~al.}(1997){Smail}, {Ivison}, \& {Blain}}]{Smail97}
{Smail}, I., {Ivison}, R.~J., \& {Blain}, A.~W. 1997, \apjl, 490, L5

\bibitem[{{Steidel} {et~al.}(2000){Steidel}, {Adelberger}, {Shapley},
  {Pettini}, {Dickinson}, \& {Giavalisco}}]{Steidel00}
{Steidel}, C.~C., {Adelberger}, K.~L., {Shapley}, A.~E., {et~al.} 2000, \apj,
  532, 170

\bibitem[{Trainor \& Steidel(2012)}]{Trainor12}
Trainor, R.~F., \& Steidel, C.~C. 2012, \apj, 752, 39

\bibitem[{Umehata {et~al.}(2017)Umehata, Tamura, Kohno, Ivison, Smail,
  Hatsukade, Nakanishi, Kato, Ikarashi, Matsuda, Fujimoto, Iono, Lee, Steidel,
  Saito, Alexander, Yun, \& Kubo}]{Umehata17}
Umehata, H., Tamura, Y., Kohno, K., {et~al.} 2017, \apj, 835, 98

\bibitem[{{Umehata} {et~al.}(2019){Umehata}, {Fumagalli}, {Smail}, {Matsuda},
  {Swinbank}, {Cantalupo}, {Sykes}, {Ivison}, {Steidel}, {Shapley}, {Vernet},
  {Yamada}, {Tamura}, {Kubo}, {Nakanishi}, {Kajisawa}, {Hatsukade}, \&
  {Kohno}}]{Umehata19}
{Umehata}, H., {Fumagalli}, M., {Smail}, I., {et~al.} 2019, Science, 366, 97

\bibitem[{Umehata {et~al.}(2021)Umehata, Smail, Steidel, Hayes, Scott,
  Swinbank, Ivison, Nagao, Kubo, Nakanishi, Matsuda, Ikarashi, Tamura, \&
  Geach}]{Umehata21}
Umehata, H., Smail, I., Steidel, C.~C., {et~al.} 2021, \apj, 918, 69

\bibitem[{{Vayner} {et~al.}(2023){Vayner}, {Zakamska}, {Sabhlok}, {Wright},
  {Armus}, {Murray}, {Walth}, \& {Ishikawa}}]{Vayner23}
{Vayner}, A., {Zakamska}, N.~L., {Sabhlok}, S., {et~al.} 2023, \mnras, 519, 961

\bibitem[{Veilleux {et~al.}(2020)Veilleux, Maiolino, Bolatto, \&
  Aalto}]{Veilleux20}
Veilleux, S., Maiolino, R., Bolatto, A.~D., \& Aalto, S. 2020, \aapr, 28

\bibitem[{Wang {et~al.}(2013)Wang, Brandt, Luo, Smail, Alexander, Danielson,
  Hodge, Karim, Lehmer, Simpson, Swinbank, Walter, Wardlow, Xue, Chapman,
  Coppin, Dannerbauer, De~Breuck, Menten, \& van~der Werf}]{Wang13}
Wang, S.~X., Brandt, W.~N., Luo, B., {et~al.} 2013, \apj, 778, 179

\bibitem[{White {et~al.}(2012)White, Myers, Ross, Schlegel, Hennawi, Shen,
  McGreer, Strauss, Bolton, Bovy, Fan, Miralda-Escude, Palanque-Delabrouille,
  Paris, Petitjean, Schneider, Viel, Weinberg, Yeche, Zehavi, Pan, Snedden,
  Bizyaev, Brewington, Brinkmann, Malanushenko, Malanushenko, Oravetz, Simmons,
  Sheldon, \& Weaver}]{White12}
White, M., Myers, A.~D., Ross, N.~P., {et~al.} 2012, \mnras, 424, 933

\bibitem[{White \& Rees(1978)}]{White78}
White, S. D.~M., \& Rees, M.~J. 1978, \mnras, 183, 341

\bibitem[{{Wong} {et~al.}(2022){Wong}, {Oudshoorn}, {Sofovich}, {Green},
  {Shah}, {Indebetouw}, {Meixner}, {Hacar}, {Nayak}, {Tokuda}, {Bolatto},
  {Chevance}, {De Marchi}, {Fukui}, {Hirschauer}, {Jameson}, {Kalari},
  {Lebouteiller}, {Looney}, {Madden}, {Onishi}, {Roman-Duval}, {Rubio}, \&
  {Tielens}}]{Wong22}
{Wong}, T., {Oudshoorn}, L., {Sofovich}, E., {et~al.} 2022, \apj, 932, 47

\bibitem[{Zhang {et~al.}(2023{\natexlab{a}})Zhang, Cai, Li, Liang, Kashikawa,
  Ma, Wu, Li, Johnson, Kikuta, Ouchi, \& Fan}]{Zhang23a}
Zhang, H., Cai, Z., Li, M., {et~al.} 2023{\natexlab{a}}, arXiv e-prints

\bibitem[{Zhang {et~al.}(2023{\natexlab{b}})Zhang, Cai, Xu, Shimakawa, Battaia,
  Prochaska, Cen, Zheng, Wu, Li, Dou, Wu, Zabludoff, Fan, Ai, Golden-Marx, Li,
  Lu, Ma, Wang, Wang, \& Yuan}]{Zhang23}
Zhang, S., Cai, Z., Xu, D., {et~al.} 2023{\natexlab{b}}, Science, 380, 494

\end{thebibliography}

\newpage
\appendix

\section{Spectroscopic Identifications of Low Redshift Objects in the Field} \label{sec:low-z}

Our deep observations of the \system\ system enabled us to acquire spectroscopic redshifts for multiple low-$z$ objects within the field-of-view. In this Appendix, we present the spectra of the four foreground objects listed in Table\,\ref{tab:objects}. 

\begin{figure*}
    \includegraphics[width=\textwidth]{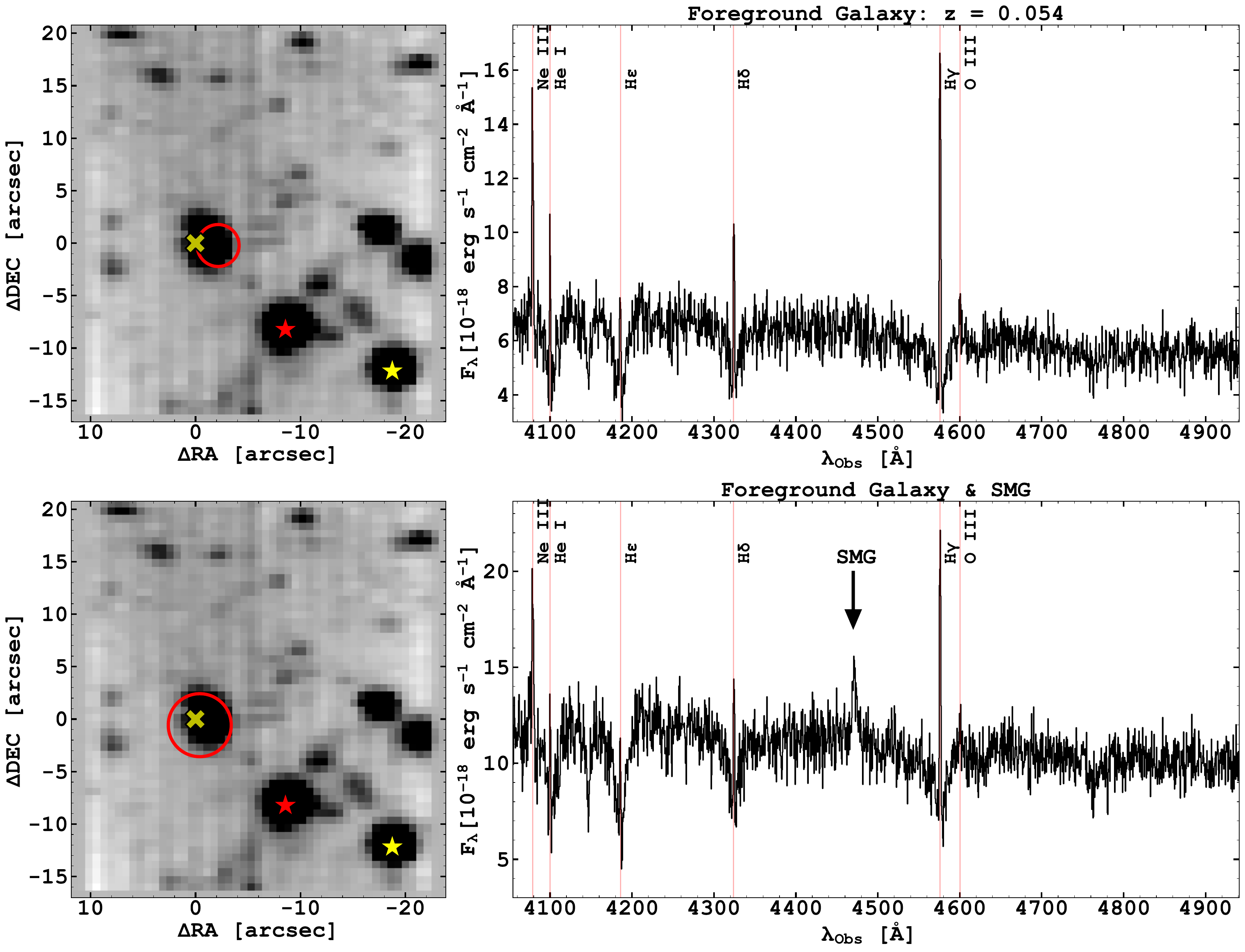}
    \caption{Spectra extracted near Obj 1 from the coadded KCWI datacube. The left column shows the extracted region by the red aperture, and the right column displays the spectrum. In the top panel, we shift the aperture away from the SMG, while we include the position of the SMG by increasing the aperture in the bottom panel. We mark emission lines from Obj 1 at $z = 0.054$, as well as the \Lya\ line from the SMG at $z=2.674$.}
    \label{fig:galaxy_blend}
\end{figure*}

Obj 1 at $z = 0.054$ is of significant interest due to its nearly overlapping position relative to the SMG (Figure\,\ref{fig:finderchart}). As a result, we need to make sure that the emission lines from the foreground galaxy do not interfere with our analysis of the \Lya\ line at the SMG redshift ($z = 2.674$). To illustrate any possible line blending, we show KCWI spectra extracted from two slightly offset aperture in Figure\,\ref{fig:galaxy_blend}. In the top panels, the aperture encloses Obj 1 but avoids the SMG; in the bottom panels, the aperture is more centered on the SMG. As expected, the \Lya\ line from the SMG shows up clearly in the bottom right panel. The SMG \Lya\ line is well separated from the H$\gamma$ and H$\delta$ lines of Obj 1. Given this result, it is deemed unnecessary to mask out Obj 1 while studying extended \Lya\ emission around the SMG, although we do need to remove the continuum from the foreground galaxy. 

\begin{figure}
    \centering
        \includegraphics[width=0.8\textwidth]{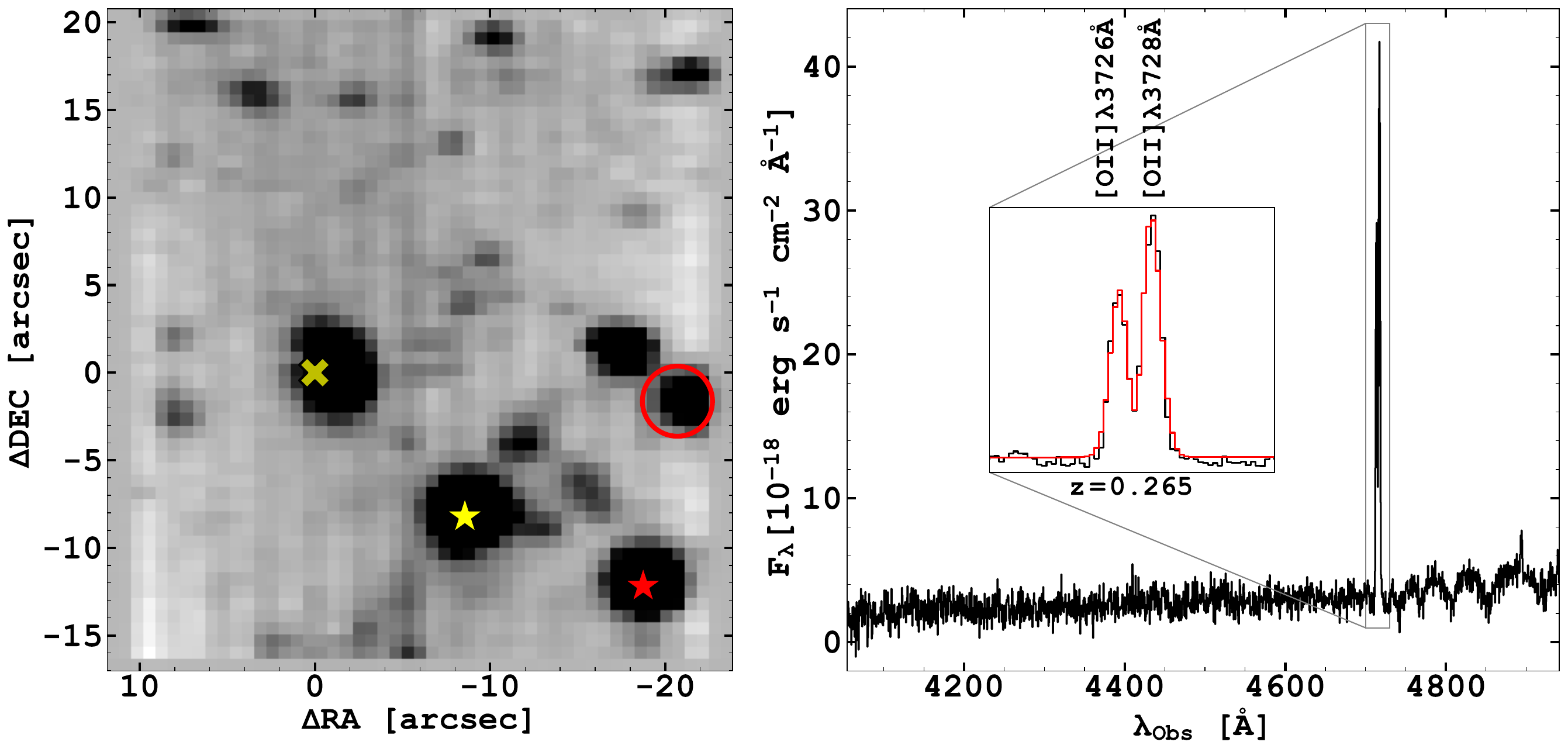}
        \includegraphics[width=0.8\textwidth]{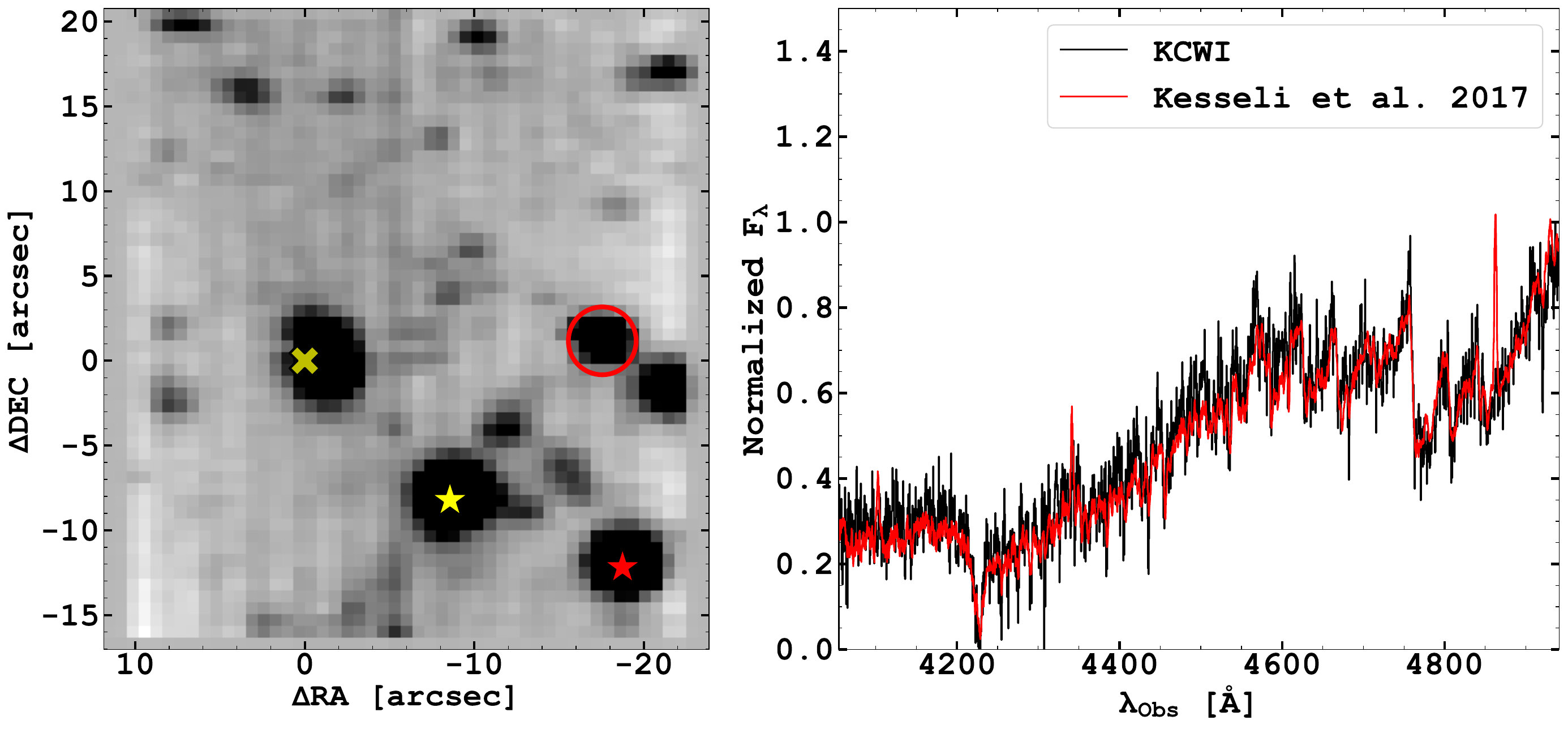}
        \includegraphics[width=0.8\textwidth]{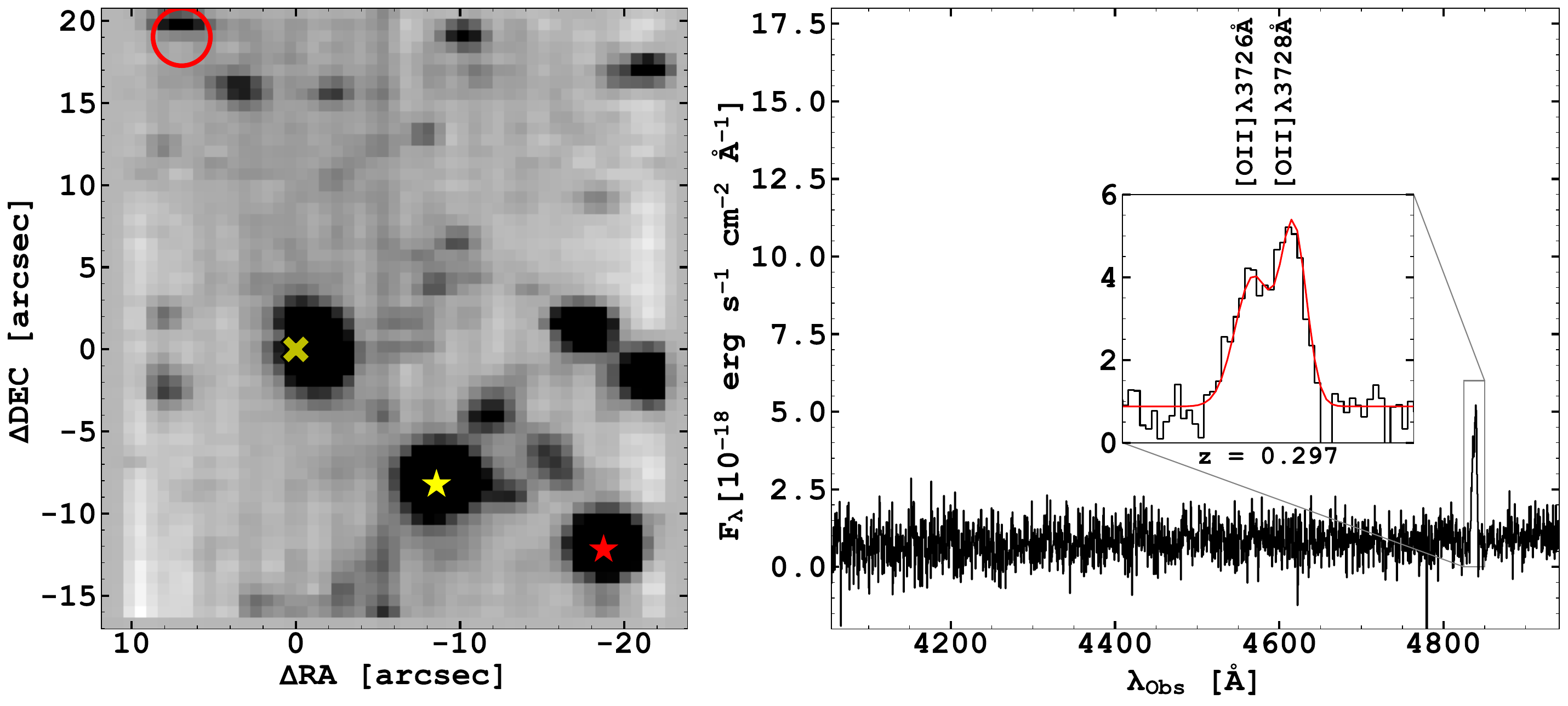}
    \caption{KCWI spectra from three foreground objects within our field-of-view. {\it Top:} Obj 2 - an \OII\ emitter at $z = 0.265$. {\it Middle:} Obj 3 - a foreground M5-type star. For comparison, we overlay a spectrum of a M5V star from \cite{Kesseli17}. {\it Bottom:} Obj 4 - another \OII\ emitter at $z = 0.297$.}
    \label{fig:foreground_targets}
\end{figure}

Figure\,\ref{fig:foreground_targets} shows the spectra from the other three foreground objects along with the positions of their extraction apertures. As the plots illustrate, we identified that Obj 2 and Obj 4 are \OII\ emitters near $z \sim 0.3$ and that Obj 3 is a M5-type star\footnote{Luminosity class is undetermined from the spectrum} in the Galaxy. 

\section{Discovery of a \Lya\ emitter at $z = 2.6918$} \label{sec:LAE2}

\begin{figure*}[!hb]
    \includegraphics[width=\textwidth]{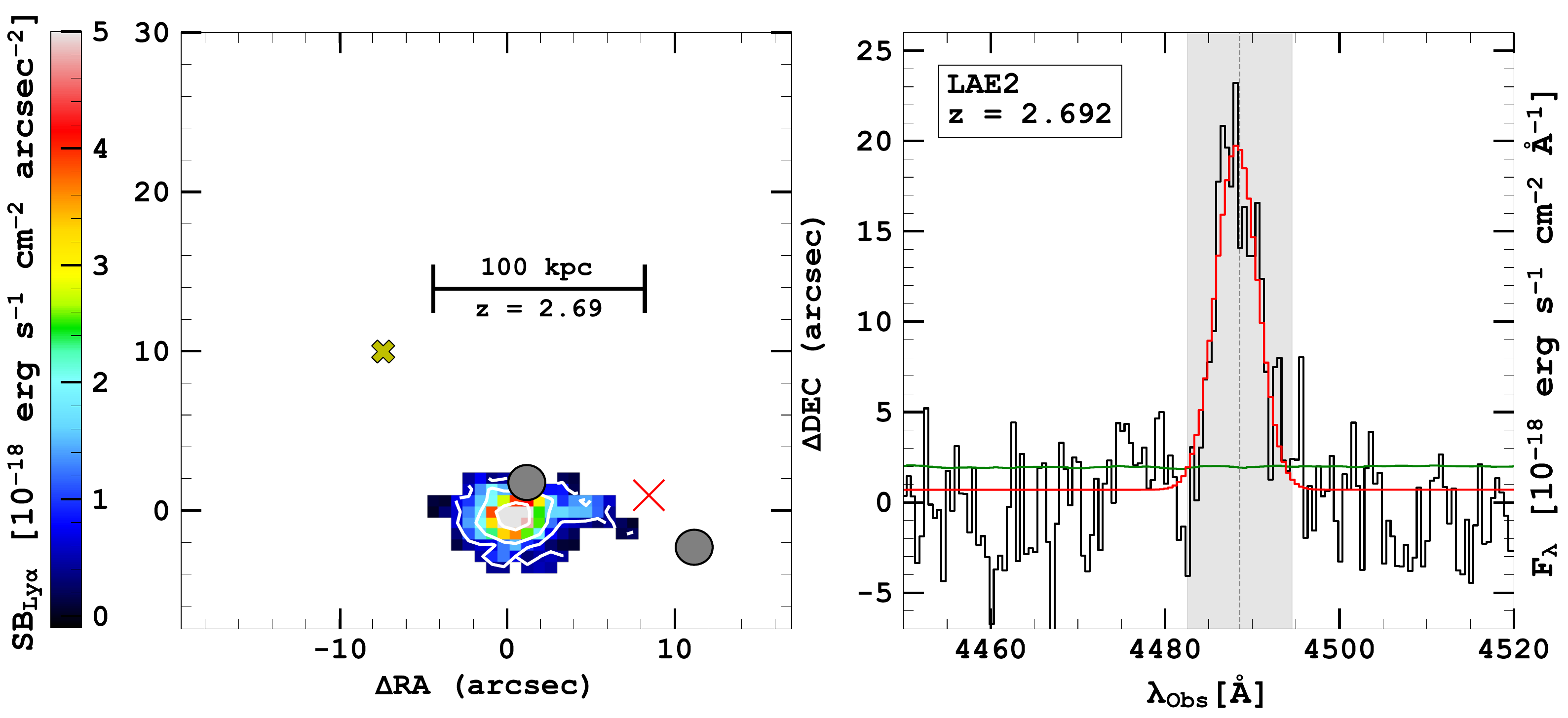}
    \caption{\Lya\ signal at $z = 2.692$. In the left panel, we show the "optimally extracted" \Lya\ surface brightness map. The position of \Chost\ is marked by the red crosshair. The white contours correspond to $\rm S/N = 2$, $5$, $10$. In the right panel, we integrate the flux within the $\rm S/N = 2$ contour and plot the spectrum as a function of the observed wavelength. The gray region corresponds to the maximum width of the 3D aperture used to generate the optimally extracted image ($\sim \pm 400$\,\kms). A 1D Gaussian is used to fit the emission line, and the red curve corresponds to the line of best fit. The green curve is the error spectrum taken from the same 3D aperture. The vertical dashed line represents the redshift of LAE2 from the observed \Lya\ line. }
    \label{fig:comp_b}
\end{figure*}

As we search the KCWI datacube for emission near our three primary targets, we detected a bright \Lya\ signal near the position of \bgQSO. We designate the \Lya\ emitter as LAE2 and show its \Lya\ surface brightness map in Figure \ref{fig:comp_b}. The total integrated \Lya\ Luminosity within the $\rm S/N = 2$ contour is $3.6 \pm 0.1 \times 10^{42}$ erg s$^{-1}$. We show the integrated spectrum in the right panel of Figure \ref{fig:comp_b} and fit its profile with a Gaussian profile. The best fit places the redshift of the \Lya\ line at $z = 2.6918 \pm 0.0006$, which is in agreement with the second redshift of the CO emitter \Chost\ at $z = 2.6917$ (Table \ref{tab:objects}). The \Lya\ emission is clearly extended, it covers a total area of $\approx 20$\,arcsec$^2$ ($\approx 1200$\,kpc$^{2}$), and the longest extent of $\approx 90$\,kpc along the EW direction.

LAE2 is at $\delta v = +1500$\,\kms\ relative to the SMG, making it unlikely to be part of its CGM. However, there is a DLA near the redshift of LAE2 along the line of sight towards \bgQSO\ (designated as subsystem C in \citetalias{Fu21}; see Figure\,\ref{fig:emission_absorption} top left panel). The DLA has two sub-components (C1 and C2) that correspond to the two CO emission redshifts of \Chost. The proximity of LAE2 to \bgQSO\ and its redshift establishes it as the most likely host galaxy of the C2 absorber. 

We now know that the DLA ``subsystem C'' has two possible hosts: LAE2 detected in \Lya\ by KCWI and \Chost\ in CO(3-2) by ALMA. The two hosts are separated by $\sim$7\arcsec\ on the sky. Interestingly, LAE2 is undetected in the deep ALMA CO datacube, and \Chost\ does not show any \Lya\ emission in the KCWI data. This case illustrates that both optical IFU observations and ALMA spectral line imaging are needed to unveil the diverse host galaxies of \Lya\ absorbers. 

\section{Overlapping \Lya\ Nebulae near \bgQSO} \label{sec:QSO1_south}

\begin{figure}[!ht]
    \hspace*{-0.5cm}      
    \centering
    \includegraphics[width=\textwidth]{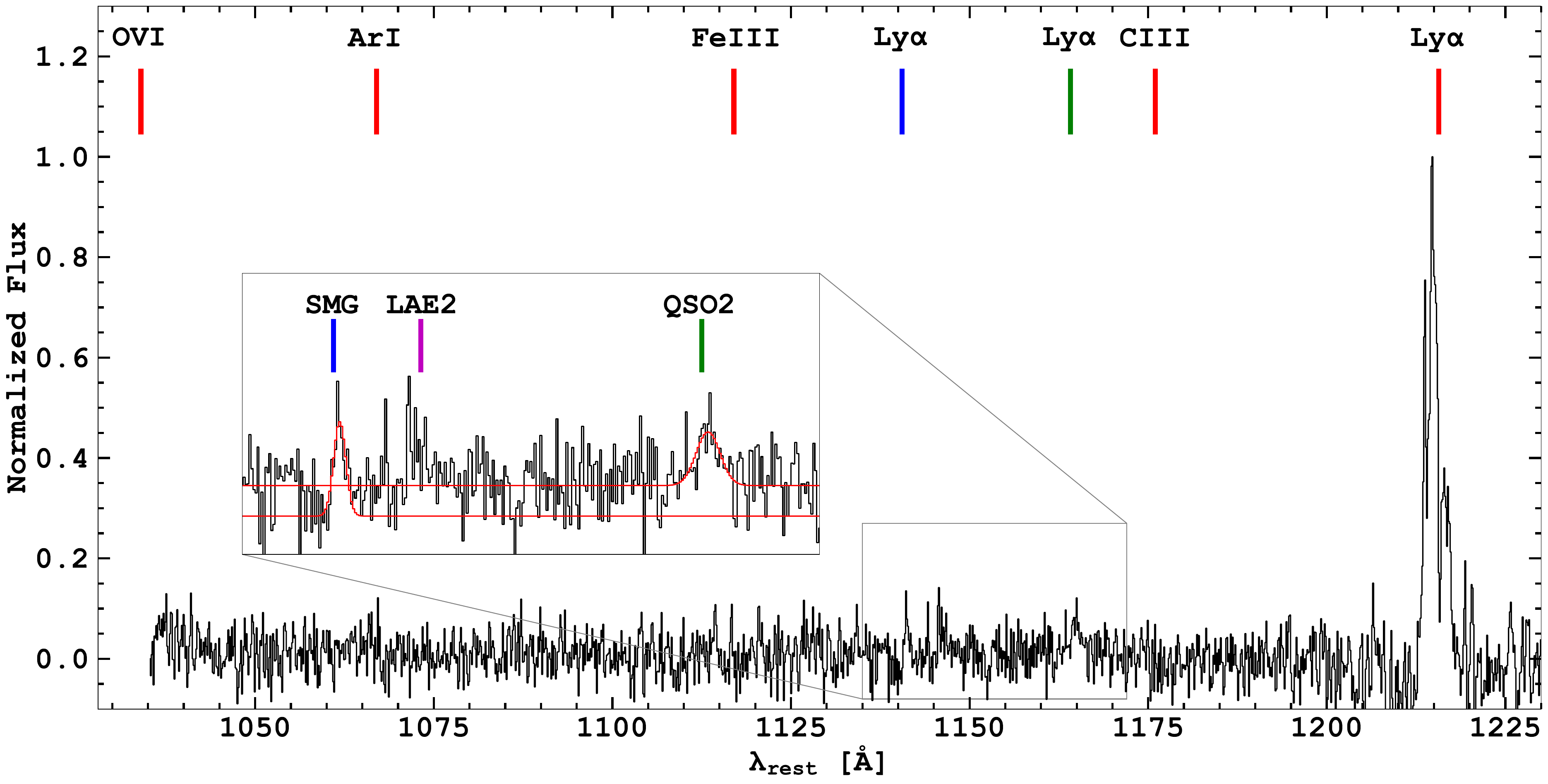}
    \caption{Full KCWI spectrum extracted from the QSO and continuum subtracted datacube with an aperture of 2\arcsec\ in radius and centered 1\farcs5 to the southwest of \bgQSO. The spectrum is plotted in the rest-frame of \bgQSO\ ($z = 2.916$). The rest-frame wavelengths of possible UV nebular emission lines are marked by red vertical lines. The \Lya\ emission lines associated with the SMG and the \fgQSO\ are marked by blue and green lines, respectively. The inset provides a closer look at the \Lya\ lines associated with the two foreground CGMs and LAE2 (purple line). No prominent emission lines near the \fgQSO\ redshift is expected to blend with the \Lya\ emission lines of the two foreground targets.}
    \label{fig:qso_extended_neb}
\end{figure}

The area around \bgQSO\ is crowded. First, there is a luminous \Lya\ nebula around \bgQSO. Further, the sightline intercepts three (sub-)DLAs at the redshifts of the SMG, the LAE2 (Appendix\,\ref{sec:LAE2}), and the \fgQSO; and it appears that the two foreground gas streams seen in \Lya\ emission also pass in front of \bgQSO. Finally, a \Lya\ emitter LAE2 appears just to the south of \bgQSO. Given that \bgQSO\ itself hosts the brightest \Lya\ nebula among the three primary galaxies, it is worth investigating whether the detected \Lya\ emission lines apparently in its foreground were actually shorter wavelength emission lines at the redshift of \bgQSO. 

In Figure\,\ref{fig:qso_extended_neb}, we display the full KCWI spectrum taken from the region of enhanced \Lya\ SB at the position of \bgQSO\ in the CGMs of the SMG and \fgQSO\ (Figure\,\ref{fig:moment_maps}). Plotted in the rest frame of \bgQSO, we indicate the wavelengths of other nebular lines between \Lyb\ and \Lya\ that could have produced a mis-identified ``\Lya'' line at lower redshifts. The composite spectra of Lyman break galaxies \citep{Shapley03,Jones12,Berry12} show an absence of emission lines in this wavelength range. So we looked into the composite QSO spectrum from the SDSS \citep{Berk01} and found four lines: C\,{\sc iii} $1176$\,\AA, Fe\,{\sc iii} $1117$\,\AA, Ar\,{\sc i} $1067$\,\AA, and O\,{\sc vi} $1034$\,\AA. As the Figure shows, none of these lines could explain the emission lines detected between 1135 and 1175\,\AA\ in the rest-frame of \bgQSO. Therefore, we conclude that it is correct to interpret them as the \Lya\ lines at the redshifts of the SMG, the \fgQSO\ redshift, and LAE2.

\section{Gas Infall/Outflow in \bgQSO?} \label{sec:QSO1_vel}

The \Lya\ nebula surrounding \bgQSO\ is not only the brightest out of the three galaxies studied here, but the moment 1 map reveals clear kinematic structure. We find that the filamentary structure, that extends nearly $\sim 150$\,kpc, is divided into two radial velocity components ($\delta v \sim \pm 300$\,\kms), and the QSO sits near the beginning of this velocity gradient. We may be observing an outflow within the CGM emanating from the QSO, which can power the observed \Lya\ nebula through shocks that form from the exiting gas. In this scenario, our sightline is edge-on to the direction of the outflow, and the $\pm \delta v$ components correspond to gas flowing towards and away from our line of sight, respectively. The large outflow may explain why \bgQSO\ has a brighter \Lya\ nebula than \fgQSO\ despite both QSOs having similar bolometric luminosities. 

Recent KCWI observations by \cite{Zhang23} of the MAMMOTH-1 \Lya\ nebula at $z=2.3$ \citep{Cai17} show remarkable similarities with \bgQSO. Their moment 1 map revealed similar kinematic structure. Specifically, the velocities are comparable with those detected in \bgQSO\ and the nebula has a similar morphology. The authors utilize other emission lines (\HeII\ and C\,{\sc iv}) to conclude that the bright nebula may be gas recycling back into the galaxy from a previous outflow event. From the \Lya\ line alone, we cannot speculate any further on the true nature of \bgQSO's nebula, but it may warrant future investigation.

\end{document}